\documentclass[aps,prb,twocolumn,superscriptaddress,amsmath,amssymb,longbibliography]{revtex4-2}

\usepackage{graphicx}
\usepackage{bm}
\usepackage[colorlinks=true,linkcolor=blue,citecolor=blue,urlcolor=blue]{hyperref}

\graphicspath{{figures/}}

\begin{document}

\title{Chiral Phonons and Giant Anisotropic Photoresponse in Quasi-1D van der Waals Semiconductor ZrSnS$_3$}

\author{Zahir Muhammad}
\thanks{Z.M. and S.M. contributed equally to this work.}
\affiliation{National Key Laboratory of Spintronics, Hangzhou International Innovation Institute, Beihang University, Hangzhou 311115, P.R. China}

\author{Shashi B. Mishra}
\email{smishra9@binghamton.edu}
\affiliation{Department of Physics, Binghamton University-SUNY, Binghamton, New York 13902, USA}

\author{Gayatri}
\affiliation{University of Warsaw, Faculty of Physics, 02-093 Warsaw, Poland}

\author{Grzegorz Krasucki}
\affiliation{University of Warsaw, Faculty of Physics, 02-093 Warsaw, Poland}

\author{Wajid Ali}
\email{w.ali2@uw.edu.pl}
\affiliation{University of Warsaw, Faculty of Physics, 02-093 Warsaw, Poland}

\author{Katarzyna Olkowska-Pucko}
\affiliation{University of Warsaw, Faculty of Physics, 02-093 Warsaw, Poland}

\author{Obaid Iqbal}
\affiliation{School of Materials Science and Engineering, Institutes of Physical Science and Information Technology, Anhui University, Hefei 230601, P.R. China}

\author{Zia Ur Rehman}
\affiliation{Nanoscale Synthesis \& Research Laboratory, Department of Applied Physics, University of Karachi, Karachi, Pakistan}

\author{Aziz Ur Rahman}
\email{aziz@ustc.edu.cn}
\affiliation{National Key Laboratory of Spintronics, Hangzhou International Innovation Institute, Beihang University, Hangzhou 311115, P.R. China}

\author{Maciej R. Molas}
\affiliation{University of Warsaw, Faculty of Physics, 02-093 Warsaw, Poland}

\author{Lin Xiaoyang}
\email{XYLin@buaa.edu.cn}
\affiliation{National Key Laboratory of Spintronics, Hangzhou International Innovation Institute, Beihang University, Hangzhou 311115, P.R. China}

\author{Weisheng Zhao}
\affiliation{National Key Laboratory of Spintronics, Hangzhou International Innovation Institute, Beihang University, Hangzhou 311115, P.R. China}

\begin{abstract}
Low-dimensional van der Waals semiconductors with reduced symmetry provide a unique platform for exploring anisotropic physical properties. The quasi-one-dimensional family MXQ$_3$ (M = Hf, Zr; X = Sn; Q = S, Se) exhibits notable structural anisotropy, where zigzag atomic chains influence optical phenomena such as birefringence. This study investigates anisotropic lattice dynamics in ZrSnS$_3$ using angle- and polarization-dependent Raman spectroscopy. Temperature-dependent measurements reveal anharmonic phonon behavior, indicating strong phonon-phonon coupling. Density functional theory calculations show good agreement with the experimentally observed Raman spectra, validating the microscopic description of the lattice dynamics. We also observe a helicity-dependent intensity and a reversal in phonon intensity between lower- and higher-frequency modes under circularly polarized light, which is characteristic of chiral phonons governed by the polarization of the Zr/Sn chains. Our first-principles analysis further shows that angular-momentum-like phonon textures can emerge away from the $\Gamma$-point near mode-hybridization and avoided-crossing regions, providing microscopic insight into the observed helicity-dependent Raman signatures. Furthermore, we fabricate an optoelectronic device from a thin ZrSnS$_3$ nanowire, demonstrating a photoresponsivity of 50~mA/W under 520~nm laser excitation (1~mW/cm$^2$). The device exhibits a pronounced, power-scalable anisotropic photoresponse with a clear preferred polarization direction. These results highlight the coupling mechanisms between polarization, lattice vibrations, and charge carriers in ZrSnS$_3$, establishing it as a promising material for polarization-sensitive optoelectronics and directional quantum transport.
\end{abstract}


\maketitle

\section{Introduction}

Low-dimensional van der Waals (vdW) materials represent a significant platform for investigating unique physical phenomena resulting from reduced dimensionality, symmetry breaking, and quantum confinement \cite{Iijima1991,Chen2019,Jotzu2014,Tian2017}. The weak interlayer coupling and atomically clean interfaces in these systems facilitate the precise engineering of physical properties, with the interplay among lattice symmetry, orbital hybridization, and electronic correlations opening substantial opportunities in spintronics, optoelectronics, and quantum materials research \cite{Guo2024,Gish2024,Sierra2021,Jiang2018,Huang2018}. Polarization-resolved studies further enrich this landscape, revealing distinct degrees of freedom such as phonon helicity \cite{Chen2015}. A key distinction arises in the optical biaxiality of certain low-symmetry crystals, which contrasts with the highly symmetric, uniaxial optical response typical of conventional 2D vdW layered crystals \cite{Arora2017,Ho2004,Kim2020}.

The recent studies of quasi-one-dimensional (1D) vdW materials have emerged as a promising frontier. These materials are characterized by anisotropic chain structures interconnected through weaker covalent, ionic, or vdW bonds in other crystallographic directions, enabling novel orientation-dependent device architectures \cite{Liu2025,Saito2025,Nataj2024,Balandin2022,Zhu2024}. Among them, group IV--V--VI ternary chalcogenides (MSnS$_3$, M = Hf, Zr) have attracted significant interest due to their low-symmetry crystal structure, semiconducting behavior, and strong structural anisotropy \cite{Wiegers1989,Meetsma1993}. Their structure consists of edge-sharing octahedral chains weakly bonded \textit{via} vdW interactions, resulting in quasi-one-dimensional electronic dispersion and pronounced polarization-dependent optical absorption. This anisotropy manifests in effective masses and birefringent responses, highlighting their potential for polarization-resolved optoelectronics \cite{Sujith2021,Shahmohamadi2021,OseiAgyemang2019,Ben2021}. For instance, HfSnS$_3$ has been identified as a strongly anisotropic \textit{p}-type semiconductor with excellent broadband photodetection capabilities \cite{Lu2023}. In contrast, its isostructural analogue, ZrSnS$_3$, remains relatively unexplored \cite{Meetsma1993}. The substitution of Hf with Zr is expected to modify atomic coordination and orbital hybridization, potentially leading to significant alterations in lattice vibrations and electronic structure, which warrants a comprehensive investigation.

Probing phonon dynamics via Raman scattering spectroscopy is an effective, non-destructive method for investigating lattice dynamics and crystallographic symmetry in such low-dimensional materials \cite{Zhang2018,Cong2020,Xu2021}. Polarized Raman spectroscopy is powerful for capturing directional anisotropy of phonon modes, elucidating bonding orientation, and revealing selection rules dictated by reduced symmetry. Furthermore, helicity-resolved Raman studies in systems like monolayer TMDs and 1D materials have been instrumental in uncovering valley-dependent selection rules and chiral phonons \cite{Arora2017,Ho2004,Kim2020,Liu2025,Saito2025,Nataj2024,Wu2022,Yin2021}. These findings underscore polarized Raman spectroscopy as a vital tool for deciphering symmetry-related vibrational phenomena in anisotropic, low-symmetry crystals like the quasi-1D vdW materials.

In this work, we synthesize ZrSnS$_3$ single crystals and investigate their structural, vibrational, and optoelectronic properties. Using angle-resolved polarized Raman spectroscopy supported by first-principles calculations, we reveal strong in-plane anisotropy in the lattice dynamics. Our helicity driven Raman measurements further uncover a pronounced circular-polarization dependent phonon response, reflecting the coupling between phonon helicity, lattice vibrations, and the anisotropic Zr/Sn chain framework. Temperature-dependent Raman spectroscopy provides additional insight into phonon anharmonicity and phonon-phonon scattering processes. In addition, devices fabricated from thin ZrSnS$_3$ nanowires exhibit an anisotropic photoresponse with a preferred polarization direction, demonstrating their potential for polarization-sensitive optoelectronics. This work provides insights into the lattice dynamics of low-symmetry quasi-1D semiconductors and highlights ZrSnS$_3$ as a potential platform for anisotropic transport, helicity-sensitive vibrational phenomena, and directional optoelectronic applications.

\section{Results and Discussion}

\begin{figure*}[t]
\centering
\includegraphics[width=\textwidth]{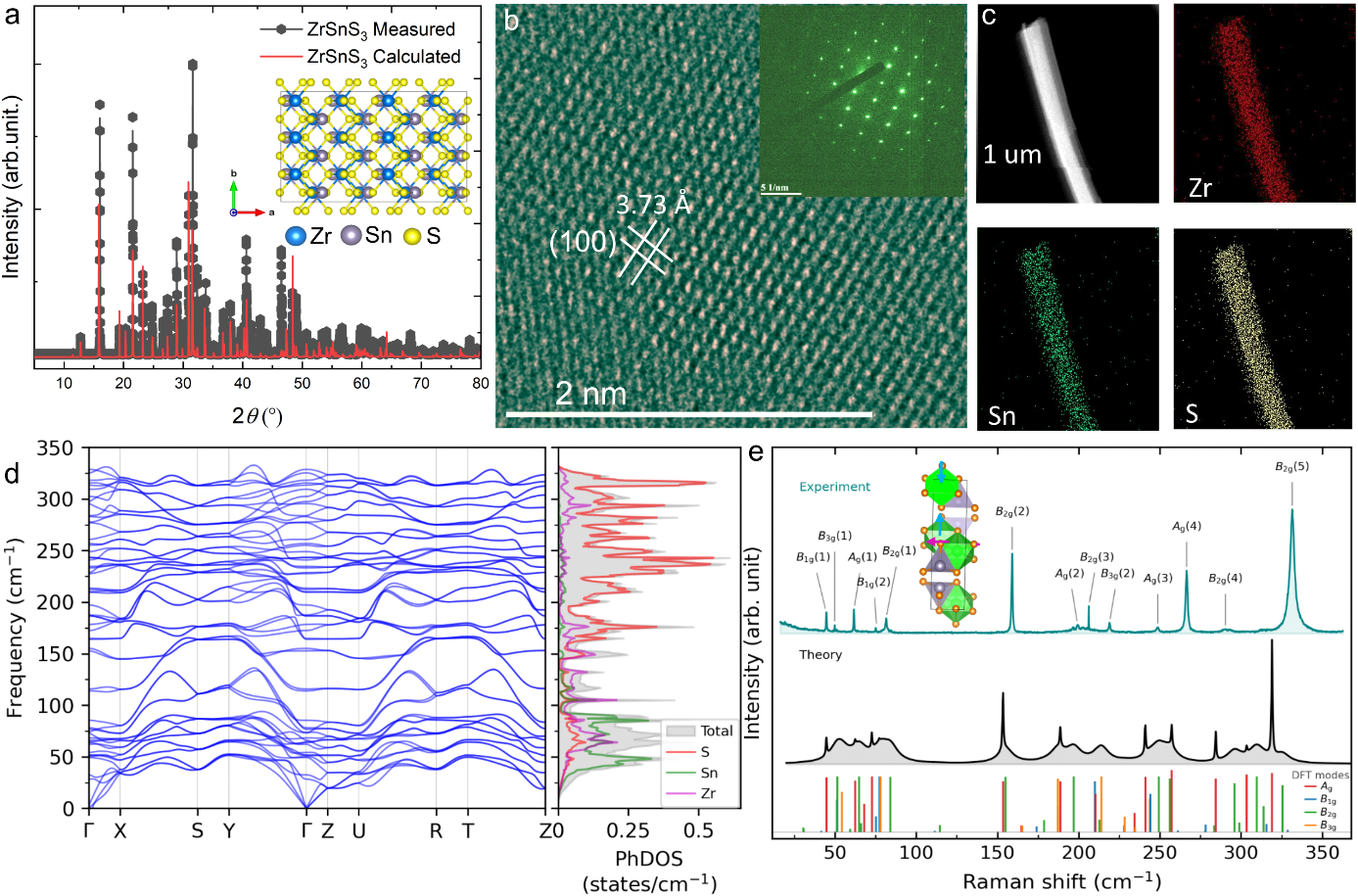}
\caption{\textbf{Structure characterization and Raman modes analysis of ZrSnS$_3$:} (a) Powder XRD analysis of grounded ZrSnS$_3$ single crystals compared with standard ZrSnS$_3$ data, inset: structure configuration of ZrSnS$_3$, (b) HAADF STEM shows the crystal structure with lattice spacing of 3.73~\AA, inset: the corresponding SAED pattern. (c) TEM image and its elemental mapping showing Zr, Sn, and S, respectively. (d) Phonon dispersion, along with total and atom-projected phonon density of states. (e) Experimental Raman spectrum measured at 5~K using 785~nm excitation and 500~$\mu$W laser power, compared with the unpolarized first-principles Raman spectrum. The theoretical spectrum is Lorentzian-broadened with FWHM $\approx 11$~cm$^{-1}$. The bottom stick spectrum shows the bare DFT mode frequencies, with intensities proportional to the calculated Raman activity and colors denoting phonon symmetry. The inset shows the structure and mode direction along the atoms for out of plane and in-plane mode.}
\label{fig1}
\end{figure*}

The high-quality wire-shaped single crystals of ZrSnS$_3$ were synthesized via the chemical vapor transport (CVT) method. The experimental details can be found in the experiments and methods part of the article. Figure~\ref{fig1}(a) reveals the powder X-ray diffraction (XRD) of the ground ZrSnS$_3$ single crystal, which was further compared and matched with the standard XRD data of ZrSnS$_3$ of JCPDS No.\ 44-1494. The calculated XRD data confirm the orthorhombic structure with a space group of $Pnma$ (62). The inset in Fig.~\ref{fig1}(a) shows the crystal structure configuration image of ZrSnS$_3$.

Scanning transmission electron microscopy (STEM) was used to measure the microstructure of ZrSnS$_3$. Figure~\ref{fig1}(b) displays a typical high-resolution STEM image of ZrSnS$_3$ with an extracted lattice atomic spacing of 3.73~\AA, which corresponds to the (100) lattice plane, confirming the atomic distance between the corresponding Zr/Sn and S atoms. The inset in Fig.~\ref{fig1}(b) indicates the selected area electron diffraction (SAED) pattern as detected from the corresponding high-resolution STEM image, which reveals the orthorhombic crystal structure of ZrSnS$_3$. The stoichiometric ratio was confirmed by measuring the energy-dispersive X-ray spectroscopy (EDS) (see Fig.~S1 in the Supplemental Material~\cite{SI}), and the uniform distribution of each element, such as Zr, Sn, and S, was further affirmed by typical TEM elemental mapping of the single crystal in Fig.~\ref{fig1}(c). The inset table in Fig.~S1 clearly revealed the stoichiometric ratio of Zr, Sn, and S of almost 1:1:3. Moreover, the elemental mapping shows the existence of Zr, Sn, and S in the as-grown crystal, which are uniformly distributed in the wire-type sample.

The lattice dynamics of ZrSnS$_3$ single crystals were studied by Raman spectroscopy (RS) and first-principles Density Functional Theory (DFT). The inset in Fig.~\ref{fig1}(a) illustrates the primitive cell and highlights the anisotropic bonding network. Figure~\ref{fig1}(d) presents the calculated phonon dispersion and atom-projected phonon density of states, with phonon frequencies extending up to 330~cm$^{-1}$. The low-frequency modes below 85~cm$^{-1}$ are dominated by the heavier Sn and Zr atoms, while the intermediate region from 85 to 200~cm$^{-1}$ contains mixed Sn, Zr, and S contributions associated with coupled bending and stretching vibrations of the Sn--S--Zr network. Above 200~cm$^{-1}$, the optical modes are mainly sulfur-derived, reflecting the lighter mass of S and stronger Sn--S bonding.

\begin{figure*}[t]
\centering
\includegraphics[width=\textwidth]{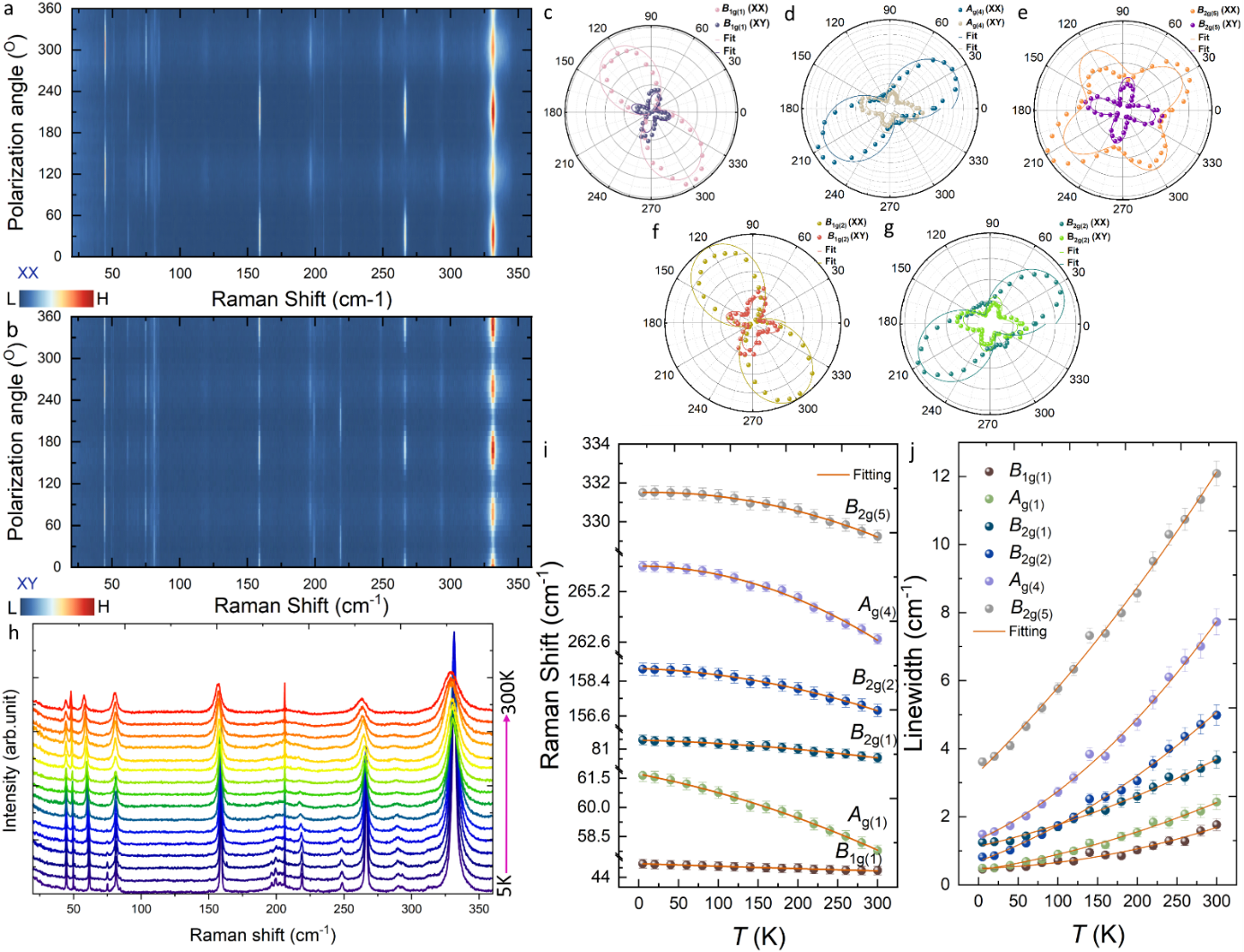}
\caption{\textbf{Polarization and temperature-dependent RS of ZrSnS$_3$:} (a, b) angle-dependent contour map of the RS using parallel and perpendicular polarization configurations, and (c-g) corresponding polar plots of higher-intensity modes at both parallel and vertical configurations, respectively. The measurement was performed at 5~K using an excitation energy of 785~nm with a laser power of 500~$\mu$W. (h) Raman spectra of ZrSnS$_3$ as a function of temperature from 5~K to room temperature; (i, j) Raman energy shift and the corresponding linewidth as a function of temperature and their fitting. The orange line shows the corresponding fit using the Klemens model. For temperature-dependent measurements, the excitation energy of 785~nm and laser power of 500~$\mu$W were used.}
\label{fig2}
\end{figure*}

Group theoretical analysis of this structure with 20-atom unit cell yields 57 optical phonon modes at the $\Gamma$-point, with irreducible representation $\Gamma = 10A_{g} + 5A_{u} + 5B_{1g} + 9B_{1u} + 10B_{2g} + 4B_{2u} + 5B_{3g} + 9B_{3u}$. Among these, 30 modes ($10A_{g}$, $5B_{1g}$, $10B_{2g}$, $5B_{3g}$) are Raman-active under $D_{2h}$ nuclear site symmetry. Experimentally, unpolarized Raman spectra collected at 5~K using 785~nm excitation reveal 13 distinct Raman peaks (see Methods), which correspond well to the calculated $A_{g}$ and $B_{g}$ symmetry vibrations involving in-plane and out-of-plane motions, as summarized in Table~\ref{tab1}. The space group $Pnma$ contains 13 Raman-active modes, while the other modes are IR-active. However, due to the backscattering arrangement of our experiment, we have only observed the $A_{g}$ and $B_{g}$ modes. The modes were further predicted from the DFT calculations and compared with the calculated modes (see Table~\ref{tab1}). The 785~nm (1.52~eV) wavelength was selected after laser-dependent studies (Fig.~S2~\cite{SI}) showed a significant ($\sim 10^{5}$ times) intensity enhancement for the strongest peak ($B_{2g}$(5)) compared to 405~nm excitation, attributed to reduced resonant absorption. Figure~\ref{fig1}(e) compares the experimental Raman spectrum at 5~K with the harmonic DFPT-calculated Raman spectrum. All 13 experimentally observed Raman modes are captured by the theoretical calculations. The calculated Raman spectrum is Lorentzian-broadened with FWHM $\approx 11$~cm$^{-1}$, while the bare DFT mode frequencies are shown as a stick spectrum to distinguish mode centers from the broadened spectral profile. The calculated peak positions agree well with the experiment, typically within $\sim$3~cm$^{-1}$, with the largest deviation of $\sim$5.5~cm$^{-1}$ observed for the higher-frequency modes. Such differences are consistent with the expected accuracy of DFT phonon calculations and can arise from finite-temperature effects, resonance conditions, the exchange-correlation functional, and anharmonic phonon renormalization beyond the present harmonic treatment \cite{Mishra2025}. Here, the DFT calculations serve as a reference for assigning the $\Gamma$-point Raman-active modes, while anharmonic effects are analyzed experimentally through the temperature-dependent Raman shifts and linewidths in Figs.~\ref{fig2}(h--j). Overall, the close agreement between theory and experiment supports the reliability of our DFT model and confirms the vibrational characteristics of ZrSnS$_3$. The phonon eigenvectors for the 13 Raman modes are shown in Fig.~S3~\cite{SI}.

\begin{table}[t]
\caption{Comparison between experimentally observed Raman-active phonon modes of ZrSnS$_3$ measured at 5~K using 785~nm excitation and DFT-calculated $\Gamma$-point phonon frequencies. The table lists the Raman mode symmetry, experimental peak positions, corresponding theoretical mode indices and frequencies, and the absolute frequency differences.}
\label{tab1}
\begin{ruledtabular}
\begin{tabular}{lcccc}
Raman mode & Experiment & Mode & Theory & Absolute difference \\
symmetry & (cm$^{-1}$) & number & (cm$^{-1}$) & (cm$^{-1}$) \\
\hline
$B_{1g}$(1) & 44.6  & 7  & 44.67  & 0.07 \\
$B_{3g}$(1) & 49.8  & 8  & 51.25  & 1.45 \\
$A_{g}$(1)  & 61.7  & 12 & 62.39  & 0.69 \\
$B_{1g}$(2) & 75.7  & 17 & 75.27  & 0.43 \\
$B_{2g}$(1) & 81.4  & 20 & 84.22  & 2.82 \\
$B_{2g}$(2) & 159.1 & 24 & 155.10 & 4.00 \\
$A_{g}$(2)  & 199.4 & 32 & 196.97 & 2.43 \\
$B_{2g}$(3) & 206.3 & 33 & 209.98 & 3.68 \\
$B_{3g}$(2) & 219.0 & 36 & 213.98 & 5.02 \\
$A_{g}$(3)  & 248.7 & 44 & 249.31 & 0.61 \\
$A_{g}$(4)  & 266.5 & 47 & 261.22 & 5.28 \\
$B_{2g}$(4) & 290.4 & 52 & 295.89 & 5.49 \\
$B_{2g}$(5) & 331.5 & 60 & 328.89 & 2.61 \\
\end{tabular}
\end{ruledtabular}
\end{table}

Angle-resolved polarized Raman spectroscopy was performed on a bulk ZrSnS$_3$ crystal across various linear polarization configurations and temperatures using 785~nm laser excitation (see Methods for details). The measurements were conducted at 5~K and spanned rotational angles from 0$^\circ$ to 360$^\circ$ of the polarization axis in both parallel (XX, $\bar{y}(zz)y$) and perpendicular (XY, $\bar{y}(zx)y$) linear polarization configurations. In the experimental arrangement, the $y$-direction corresponds to the $a$-axis of the crystal, while the $z$- and $x$-polarization directions lie in the $bc$ plane of the investigated sample.

For the orthorhombic ZrSnS$_3$ crystal with $D_{2h}$ space group, the intensity of each mode depends on the incident and scattered light polarization vectors. This relationship is described by the specific $3\times 3$ Raman tensor for each irreducible representation ($A_{g}$ and $B_{g}$ modes) in the back-scattering geometry, as outlined in Ref.~\cite{Wang2017}:
\begin{equation}
\begin{aligned}
A_{g} &= \begin{pmatrix} a & 0 & 0 \\ 0 & b & 0 \\ 0 & 0 & c \end{pmatrix}, &
B_{1g} &= \begin{pmatrix} 0 & d & 0 \\ d & 0 & 0 \\ 0 & 0 & 0 \end{pmatrix}, \\[4pt]
B_{2g} &= \begin{pmatrix} 0 & 0 & e \\ 0 & 0 & 0 \\ e & 0 & 0 \end{pmatrix}, &
B_{3g} &= \begin{pmatrix} 0 & 0 & 0 \\ 0 & 0 & f \\ 0 & f & 0 \end{pmatrix}.
\end{aligned}
\label{eq1}
\end{equation}

The Raman intensity ($I$) is proportional to the square of the projection of the polarization vectors of the incident and scattered light onto the Raman tensor, which can be written as: $I \propto \left| \hat{e}_{i} \cdot \mathfrak{R} \cdot \hat{e}_{s} \right|^{2}$, where $\hat{e}_{i}$ and $\hat{e}_{s}$ are the unit polarization vectors of the incident and scattered light, respectively, and $\mathfrak{R}$ is the Raman tensor. The detailed Raman tensor analysis can be found for the three Raman modes in the Supplemental Material~\cite{SI}. The linear polarization dependence of the Raman intensities can be fitted using the following equations \cite{Ribeiro2015}:
\begin{multline}
I^{\parallel}(\theta) = \big( |A|\sin^{2}(\theta - \phi) \\
+ |C|\cos(\alpha)\cos^{2}(\theta - \phi) \big)^{2} \\
+ |C|^{2}\sin^{2}(\alpha)\cos^{4}(\theta - \phi),
\label{eq2}
\end{multline}
\begin{multline}
I^{\bot}(\theta) = \left[ \left( |A| - |C|\cos^{2}(\alpha) \right)^{2} + |C|^{2}\sin^{2}(\alpha) \right]^{2} \\
\times \sin^{2}(\theta - \phi)\cos^{2}(\theta - \phi).
\label{eq3}
\end{multline}

These fitting equations accurately describe the intensity variations observed in the experimental data [see Figs.~\ref{fig2}(c-g)]. The angle $\theta$ represents the angle between the crystal $a$-axis and the polarization direction of the incident light ($ab$-axis). Whereas $\alpha$ and $\phi$ represent the analyzer angle (scattered polarization) and fixed sample rotational angle of the crystal, respectively. The in-plane spatial dependence of the phonon modes is evident, and the intensity variation of different Raman modes is displayed in the polar plots in Figs.~\ref{fig2}(c-g). These plots illustrate the polarization-dependent Raman response for the various vibrational modes. For the $\hat{e}_{i} \parallel \hat{e}_{s}$ configuration, four modes [$B_{1g}$(1), $B_{1g}$(2), $B_{2g}$(2), and $A_{g}$(4)] exhibit predominantly twofold symmetry, while the $B_{2g}$(5) mode shows a characteristic fourfold symmetry. For the $\hat{e}_{i} \perp \hat{e}_{s}$ configuration, all these modes demonstrate fourfold symmetry [see Figs.~\ref{fig2}(c-g)].

The solid lines in the polar plots represent the fits obtained using the above equations. The agreement between the experimental data and the fitting formulas confirms the precision of our calculations. Although the measured angular dependences generally follow the symmetry-derived Raman selection rules, small deviations between the experimental data and the idealized symmetry patterns are observed, for example in Fig.~\ref{fig2}(g). Such deviations may arise from sample inhomogeneity, slight experimental misalignment, finite polarization purity, and intrinsic effects not included in the simplified model, such as electron--phonon coupling or mode mixing. In addition, weaker Raman modes are associated with larger uncertainty in the extracted intensities, leading to increased scatter in the experimental angular profiles. Nevertheless, the overall symmetry and linear polarization dependence of the Raman modes are well captured by the fitting formulas, demonstrating the intrinsic relationship between phonon vibrations and the crystalline phase and orientation.

The RS of ZrSnS$_3$ was further studied in a temperature range of 5~K to 300~K to investigate the phonon anharmonicity and phonon-phonon interactions, as illustrated in Figs.~\ref{fig2}(h-j). As the temperature decreases, all phonon peaks display a shift towards higher frequencies as the temperature decreases from 300~K to 5~K, as illustrated in Fig.~\ref{fig2}(h). This shift indicates that the Raman modes soften as the temperature increases, a phenomenon commonly linked to anharmonic effects in the lattice \cite{Tian2017b,Balkanski1983}. These temperature-dependent variations in both the phonon frequencies and linewidths are consistent with previous findings on lattice anharmonicity in one-dimensional (1D) materials, where such effects are primarily attributed to phonon-phonon coupling, especially along the zigzag direction of the crystal structure \cite{Tian2017b,Balkanski1983}.

\begin{figure*}[t]
\centering
\includegraphics[width=\textwidth]{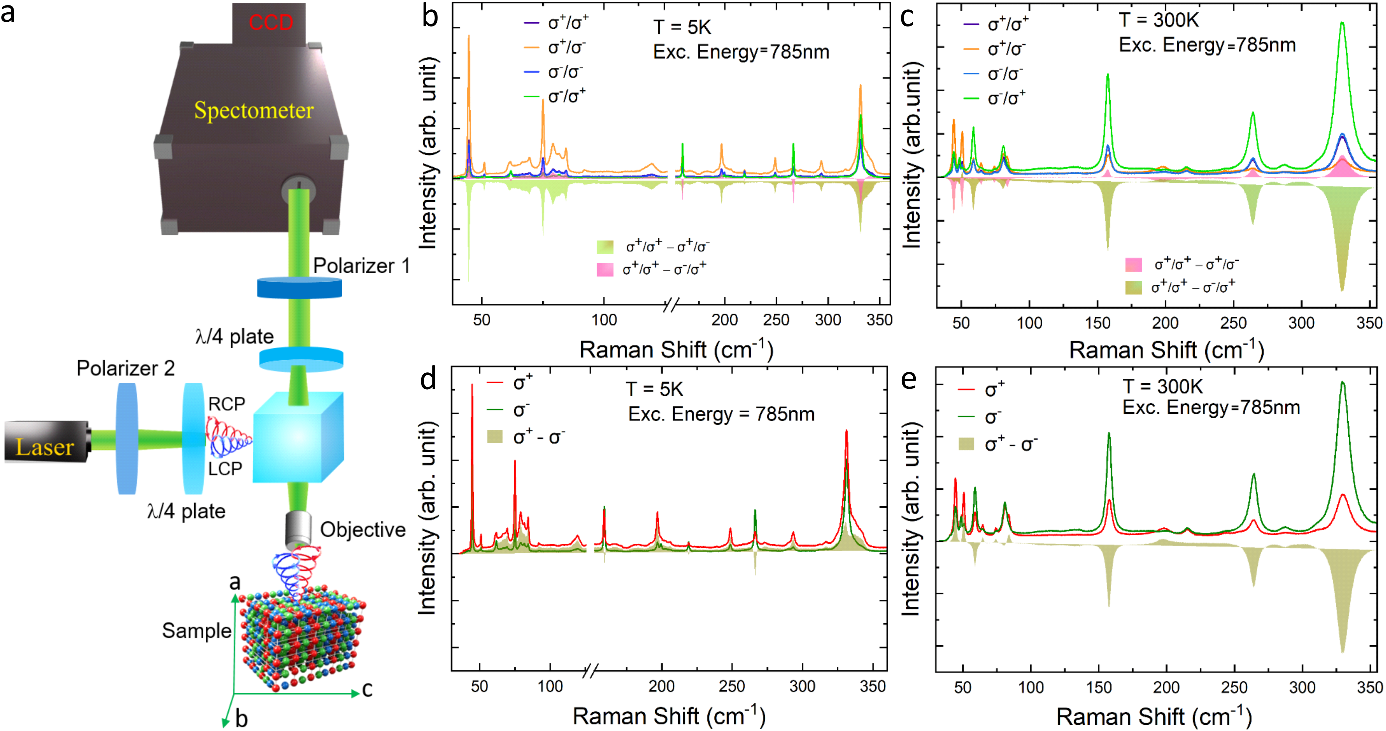}
\caption{\textbf{Circular polarized Raman spectra:} (a) Schematic of the optical setup for helicity-resolved Raman spectroscopy measurements. Circularly and helicity-resolved Raman spectra of ZrSnS$_3$ using laser excitation at 785~nm and a laser power of 1.2~mW measured at (b) 5~K and (c) 300~K. (d) \& (e) Helicity-resolved Raman spectra of ZrSnS$_3$ with left and right circular polarization ($\sigma^{+}$ and $\sigma^{-}$) and the corresponding difference shows the chirality measured at 5~K and 300~K with fixed laser power of 1.2~mW, respectively.}
\label{fig3}
\end{figure*}

The Raman shifts and linewidths of the observed phonon peaks were fitted as a function of temperature using a Voigt function. The phonon energy shift increases linearly with temperature, as seen in Fig.~\ref{fig2}(i). Specifically, the $A_{g}$(1) mode exhibits the largest shift of approximately 1.35\%, while the $B_{1g}$(1) mode shows a minimal shift of 0.13\%. These shifts strongly suggest the presence of phonon anharmonicity and phonon-phonon coupling within the crystal, as expected from the general trend of lattice anharmonicity in 1D systems \cite{Tian2017b}. The linewidths, shown in Fig.~\ref{fig2}(j), also exhibit a clear temperature dependence. This broadening of the phonon peaks is attributed to anharmonic phonon decay, where phonon-phonon interactions increase with rising temperature, resulting in enhanced phonon scattering and broader linewidths. The observed linewidth behavior further supports the notion of anharmonic decay, which is expected for 1D materials where lattice expansion and anharmonic interactions dominate the phonon scattering dynamics. To model these temperature-dependent shifts and linewidths more accurately, the Klemens model \cite{Balkanski1983} was employed to account for the anharmonicity of the phonons in ZrSnS$_3$:
\begin{equation}
\omega(T) = \omega_{0} + A\left[ 1 + \frac{2}{e^{x} - 1} \right] + B\left[ 1 + \frac{3}{e^{y} - 1} + \frac{3}{\left( e^{y} - 1 \right)^{2}} \right]
\label{eq4}
\end{equation}
and
\begin{equation}
\Gamma(T) = \Gamma_{0} + C\left[ 1 + \frac{2}{e^{x} - 1} \right] + D\left[ 1 + \frac{3}{e^{y} - 1} + \frac{3}{\left( e^{y} - 1 \right)^{2}} \right],
\label{eq5}
\end{equation}
here $\omega_{0}$ and $\Gamma_{0}+C+D$ are the harmonic frequency and linewidth at zero temperature for the optical modes, respectively; $x = \hbar\omega/(2k_{\mathrm{B}}T)$, $y = \hbar\omega/(3k_{\mathrm{B}}T)$, $\omega$ is the phonon frequency, $\hbar$ is the reduced Planck constant, and $k_{\mathrm{B}}$ is the Boltzmann constant. $A$ and $C$ are the anharmonic coefficients for three-phonon processes, while $B$ and $D$ are those for four-phonon processes. Due to this anharmonic trend, the data were fully fitted in Figs.~\ref{fig2}(i) and (j) using Eqs.~(\ref{eq4}) and (\ref{eq5}) along with a simple Voigt function to extract the fitting parameters, as seen in Table~S1~\cite{SI}. The orange line represents the fitting functions, which are well-aligned with our experimental data. This validates the presence of anharmonic phonon scattering and confirms that the phonon modes in ZrSnS$_3$ exhibit the expected behavior of phonon-phonon interactions with temperature.

However, not all modes follow the predictions of the Klemens model. Notably, the $B_{2g}$(2), $A_{g}$(4), and $B_{2g}$(5) modes, shown in Fig.~\ref{fig2}(j), exhibit a non-linear trend in the temperature range of approximately 130 to 160~K, where these modes show deviations from the expected anharmonic behavior. This behavior may be attributed to other structural factors within ZrSnS$_3$, such as phonon-phonon interactions, phonon anharmonicity, or possible anisotropies in the crystal structure, which are not fully described by the Klemens model alone. Similar behavior has been observed in other 1D materials, where additional complexities in the phonon scattering dynamics lead to deviations from conventional anharmonic models \cite{Bonini2007,Menendez1984}. Despite these deviations, the majority of the Raman modes are well-fitted by the Klemens model, further corroborating that the Raman spectra of ZrSnS$_3$ predominantly exhibit anharmonic phonon-phonon scattering.

In Fig.~\ref{fig3}, we present a detailed investigation of the circular polarization-resolved Raman spectra of ZrSnS$_3$, measured at both 5~K and 300~K using 785~nm laser excitation. Circular polarization-sensitive Raman measurements were performed using an experimental configuration composed of fixed linear polarizers in both the excitation and detection paths, combined with rotatable $\lambda/4$ wave plates to access different polarization states [see Fig.~\ref{fig3}(a)]. Two types of measurements were carried out. The results presented in Figs.~\ref{fig3}(b) and (c) were obtained for all four circular helicity combinations ($\sigma^{+}/\sigma^{+}$, $\sigma^{+}/\sigma^{-}$, $\sigma^{-}/\sigma^{+}$, and $\sigma^{-}/\sigma^{-}$), providing a comprehensive view of the helicity-dependent light--matter interactions in this quasi-1D semiconductor. In contrast, Figs.~\ref{fig3}(d) and (e) show the spectra obtained under different circular excitation conditions ($\sigma^{+}$ and $\sigma^{-}$), while the detection remained unpolarized. In this case, the spectra measured for the same excitation helicity and opposite detection helicities were summed together, enabling direct access to the chiral response of the material. The measurements were performed at both low temperature (5~K) and room temperature (300~K), allowing us to examine the temperature dependence of the chiral phonon response in ZrSnS$_3$. As helicity-resolved Raman measurements are highly sensitive to experimental artifacts, we performed control measurements on bilayer MoTe$_2$, which is not expected to host chiral phonons, as presented in Fig.~S4~\cite{SI}.

At 5~K, the spectra exhibit a pronounced circular polarization dependence. The phonon intensities observed in the helicity-conserving scattering configurations ($\sigma^{+}/\sigma^{+}$ and $\sigma^{-}/\sigma^{-}$) are nearly identical, indicating that these modes are largely insensitive to the helicity of the incident light. In contrast, the helicity-flip scattering configurations ($\sigma^{+}/\sigma^{-}$ and $\sigma^{-}/\sigma^{+}$), display a substantial enhancement of the phonon intensities. Such behavior provides direct insight into the helicity-dependent properties of phonons in ZrSnS$_3$. Several phonon modes, including $A_{g}$(1) (61.97~cm$^{-1}$), $B_{2g}$(2) (159.09~cm$^{-1}$), $B_{2g}$(3) (206.6~cm$^{-1}$), $A_{g}$(4) (266.16~cm$^{-1}$), and $B_{2g}$(5) (331.32~cm$^{-1}$), exhibit pronounced polarization-dependent intensities. In particular, the low-frequency modes are more intense in the $\sigma^{+}/\sigma^{-}$ configuration than in $\sigma^{-}/\sigma^{+}$, whereas the opposite trend is observed for the high-frequency modes [see Fig.~\ref{fig3}(b)]. This contrasting behavior reveals a clear helicity dependence of the phonon response.

At room temperature (300~K), the Raman spectra exhibit a pronounced temperature dependence. The low-frequency phonon modes that are clearly resolved at 5~K are no longer detectable at 300~K, most likely due to thermal broadening and the loss of coherence of the low-energy vibrational modes [Fig.~\ref{fig3}(c)]. In contrast, the high-frequency phonon modes remain visible and continue to display a clear polarization dependence, together with distinct intensity differences between the circular polarization configurations. These observations demonstrate that the helicity-dependent phonon response persists up to room temperature.

To further verify the presence of chiral phonons in ZrSnS$_3$, we compare the Raman spectra measured at low temperature (5~K) and room temperature (300~K) under opposite circular excitation conditions ($\sigma^{+}$ and $\sigma^{-}$), while maintaining unpolarized detection [see Fig.~\ref{fig3}(d, e)]. As shown in Fig.~\ref{fig3}(d), most phonon modes exhibit significantly higher intensities under $\sigma^{+}$ excitation than under $\sigma^{-}$ excitation. In contrast, the $B_{2g}$(2) (159.09~cm$^{-1}$) and $A_{g}$(4) (266.16~cm$^{-1}$) modes display the opposite behavior. At 300~K, the relative intensities of the spectra measured under $\sigma^{+}$ and $\sigma^{-}$ excitation change substantially, with the majority of phonon modes becoming more intense under $\sigma^{-}$ excitation than under $\sigma^{+}$ excitation. Only the $B_{1g}$(1) (44.6~cm$^{-1}$) and $B_{3g}$(1) (49.8~cm$^{-1}$) modes retain the opposite tendency. The pronounced intensity differences observed at both temperatures result in a clear non-zero differential signal for the corresponding phonon modes, providing strong evidence for the chiral nature of the phonon response in ZrSnS$_3$. Moreover, the persistence of the helicity-dependent response upon increasing the temperature from 5~K to 300~K demonstrates the robustness of the chiral phonon modes, despite the thermal suppression of certain low-energy vibrations. Notably, Fig.~\ref{fig3}(d) reveals a pronounced non-zero differential signal across the entire spectral range, including the regions between the phonon peaks, suggesting that the chiral optical response extends beyond the discrete vibrational modes and involves the electronic states of the material. Collectively, these results demonstrate a robust net chiral response in ZrSnS$_3$.

We emphasize that first-order Raman scattering primarily probes zone-center phonons; therefore, the helicity-resolved spectra should not be interpreted as a direct measurement of a single finite-$\mathbf{q}$ phonon angular momentum. Instead, the observed intensity modulation is interpreted as a helicity-dependent Raman response. Such helicity sensitivity can arise from interference between Raman tensor components and from weak relaxation of ideal selection rules caused by local symmetry breaking, mode mixing, surface effects, focusing-induced angular spread, strain, or defects \cite{Parlak2023,Kumar2024}. Similar selection-rule relaxation has been reported in circularly polarized Raman measurements of chiral tellurium, where the $\Gamma_{1}$ mode is symmetry-forbidden in the RL and LR channels but is nevertheless observed experimentally due to non-parallel incidence, finite focusing angle, and surface roughness \cite{Ishito2023}. In this regime, the Raman response can acquire sensitivity to near-$\Gamma$ and mixed phonon states, even when the ideal $\Gamma$-point angular-momentum expectation value is zero. The results demonstrate that the chiral phonons are sensitive to both the helicity and the nature of the detected polarization, evolving continuously from circular to elliptical to linear polarization and back. For same-helicity scattering ($\sigma^{+}/\sigma^{+}$), the Raman intensity of certain modes is enhanced, whereas different modes exhibit increased intensity under opposite-helicity scattering ($\sigma^{+}/\sigma^{-}$). This angular dependence provides direct evidence of phonon chirality in ZrSnS$_3$, manifested as a clear modulation of Raman intensity with polarization angle. Moreover, the observed variations in both intensity and phase as a function of $\theta$ highlight the polarization-selective nature of the Raman scattering process, underscoring the strong coupling between chiral phonons and circularly polarized light.

\begin{figure*}[t]
\centering
\includegraphics[width=\textwidth]{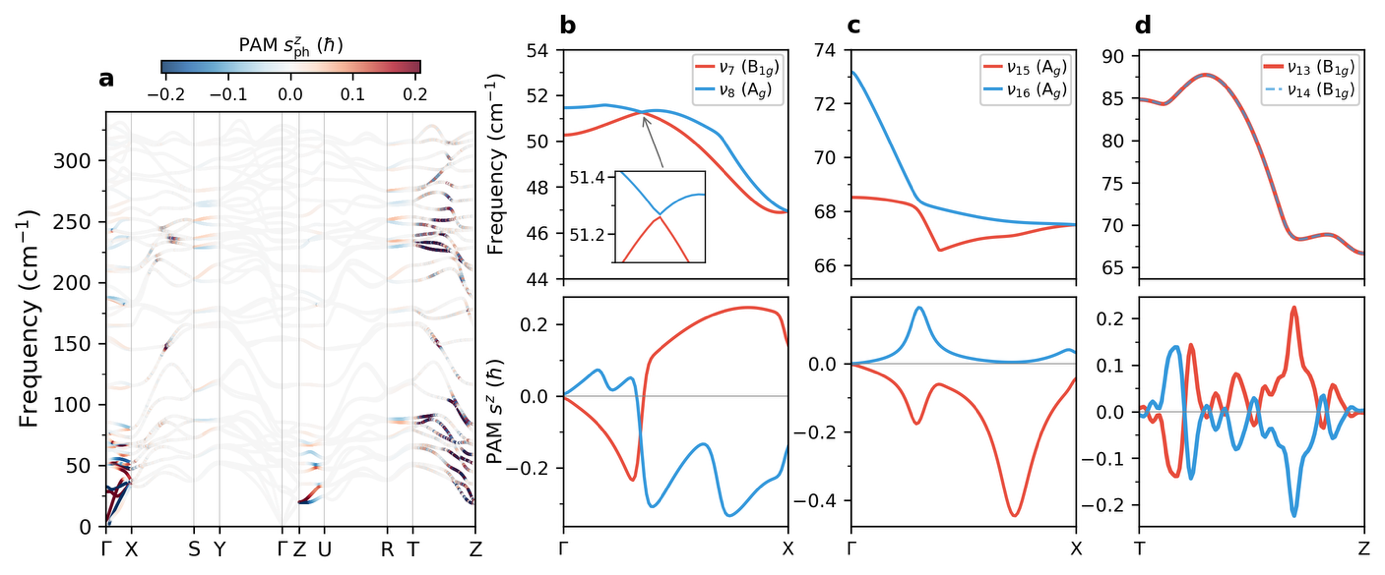}
\caption{\textbf{Chiral Phonons in ZrSnS$_3$ from theory.} (a) Calculated Phonon pseudo angular momentum (PAM) $s^{z}$, along high-symmetry directions of the Brillouin zone, showing finite chirality mainly along $\Gamma-X$, $Z-U$, and $T-Z$. (b) Avoided crossing between $B_{1g}$(1) and $A_{g}$(1) branches ($\nu = 7, 8$) along $\Gamma-X$, together with the corresponding $s^{z}$, showing a sign reversal as the modes exchange character. The zoom-in inset resolves the narrow gap at the avoided crossing. (c) Avoided crossing between $A_{g}$(2)-derived branches ($\nu = 15, 16$) along $\Gamma-X$, producing the strongest low-frequency single branch PAM response. (d) Nearly degenerate $B_{1g}$(2) phonon pairs ($\nu = 13, 14$) along $\Gamma-Z$, with opposite $s^{z}$ carried by the two partners, consistent with symmetry-enforced band sticking. Additional band-sticking features are summarized in Fig.~S7~\cite{SI}.}
\label{fig4}
\end{figure*}

These helicity-resolved Raman measurements reveal a pronounced circular-polarization-dependent Raman response in ZrSnS$_3$. Since first-order Raman scattering primarily probes $\Gamma$-point phonons, this response is best understood as a helicity-sensitive Raman signature rather than a direct measurement of a single finite-$\mathbf{q}$ phonon angular momentum. In the ideal centrosymmetric $Pnma$/$D_{2h}$ structure, which provides the baseline for assigning the $A_{g}$, $B_{1g}$, $B_{2g}$, and $B_{3g}$ Raman modes using standard Raman tensors, all $\Gamma$-point Raman-active modes have zero angular-momentum expectation value, as summarized in Table~S2~\cite{SI}. Unlike structurally chiral crystals, where phonon chirality can be enforced by crystallographic handedness or symmetry-protected degeneracies \cite{Coh2023,Kumar2026,Zhang2015,Zhu2018,Hamada2018,Ishito2023}, ZrSnS$_3$ is centrosymmetric in its average structure. Thus, the observed helicity-dependent Raman response is not attributed to intrinsic $\Gamma$-point phonon angular momentum but is instead consistent with Raman tensor interference and weak relaxation of the ideal selection rules. When inversion symmetry ($\mathcal{P}$) is weakly broken by the Sn-dominated polar perturbation discussed below, $\mathcal{PT}$ symmetry is also broken while time-reversal symmetry ($\mathcal{T}$) is retained. Under such conditions, finite phonon angular momentum can emerge from asymmetry in the force-constant matrix \cite{Ishito2023HgS,Moseni2022}. Finite phonon circular polarization may also arise through the breaking of specific mirror symmetries, even when inversion and time-reversal symmetries are nominally preserved \cite{Rao2025}.

To identify a microscopic lattice-dynamical mechanism compatible with this helicity-sensitive response, we calculated the circular-polarization character of phonon eigenvectors from first principles (see Methods). This quantity, commonly referred to as phonon pseudo-angular momentum (PAM), is used here as a continuous expectation-like measure of the imbalance between right- and left-handed circular components of the phonon eigenvectors, rather than as a discrete symmetry-protected quantum number \cite{Yang2024}. DFPT calculations for the centrosymmetric reference identify a soft polar ($B_{2u}$) mode at $\Gamma$, dominated by coherent displacement of the Sn sublattice along the $\mathbf{b}$-direction (Table~S3~\cite{SI}). Following this eigenvector produces a small inversion-breaking distortion, [$\Delta y$ (Sn) $= +0.003$] in fractional coordinates, corresponding to approximately 0.011~\AA, with Zr and S remaining nearly unchanged. The distorted structure (Table~S4~\cite{SI}) is nearly degenerate with the centrosymmetric reference, with an energy difference of about 0.21~meV per formula unit, indicating a flat polar coordinate rather than a strongly stabilized bulk polar phase. This distortion is not used to fit the Raman spectra or alter the conventional linear-polarization mode assignments; instead, it provides a possible microscopic route by which local strain, defects, surfaces, or structural fluctuations may weakly relax the ideal centrosymmetric selection rules.

\begin{figure*}[t]
\centering
\includegraphics[width=\textwidth]{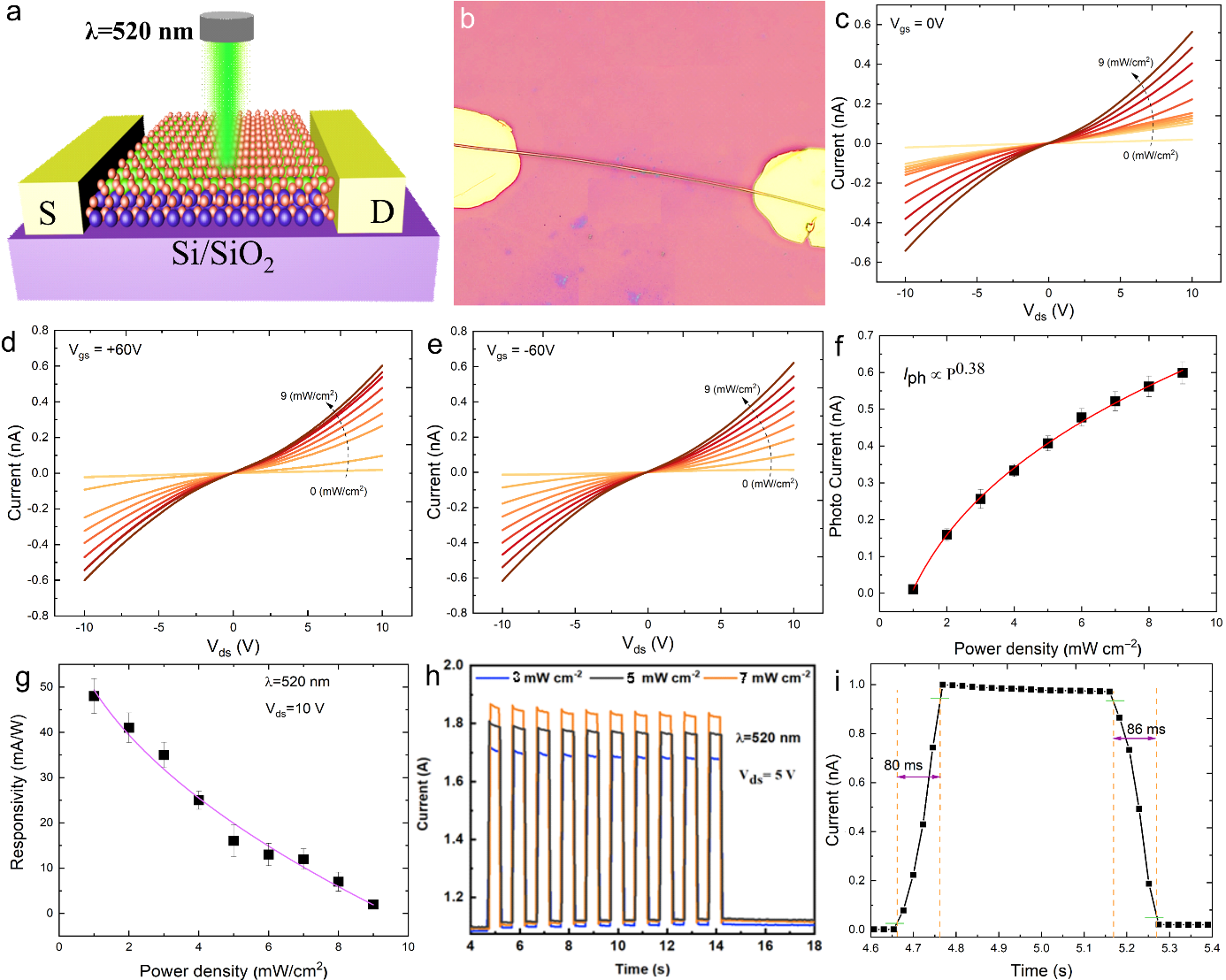}
\caption{\textbf{Performance of the photodetector based on ZrSnS$_3$ nanowire.} (a) Schematic diagram of the ZrSnS$_3$ photodetector on a SiO$_2$/Si substrate. (b) Optical microscope photograph of the ZrSnS$_3$ photodetector. Scale bar: 10~$\mu$m. (c-e) The photo current ($I_{\mathrm{ph}}$) at various power intensities of the incident 520~nm laser at gated voltage $V_{\mathrm{g}}$ ($\pm 60$~V, 0~V). (f) Power density-dependent photocurrent under 520~nm illumination and the corresponding power-law fit. (g) Responsivity of the photodetector at 10~V source under 520~nm laser illumination at various intensities. (h) Long-time response stability of the photodetector. (i) The response time of the photodetector.}
\label{fig5}
\end{figure*}

The calculated PAM distribution, shown in Fig.~\ref{fig4}(a), is highly anisotropic in momentum space, with finite values mainly along $\Gamma$--X, Z--U, and T--Z. These finite-$\mathbf{q}$ calculations therefore provide a microscopic map of where angular-momentum-like phonon textures can emerge under weakly relaxed symmetry conditions. The mode-resolved heatmap in Fig.~S6~\cite{SI} further shows that enhanced PAM is associated with two main mechanisms: mode hybridization at avoided crossings and symmetry-enforced band sticking. Along $\Gamma$--X, the avoided crossing between the $B_{1g}$(1) and $A_{g}$(1)-derived branches ($\nu = 7, 8$) redistributes circular-polarization character between the two modes [Fig.~\ref{fig4}(b)]. Although these labels refer to the parent $Pnma$ structure, the weak polar $Pna2_{1}$ distortion lowers the symmetry from $D_{2h}$ to $C_{2v}$, under which both parent $A_{g}$ and $B_{1g}$ modes transform as the same daughter $A_{1}$ representation. They can therefore hybridize along $\Gamma$--X. The zoomed inset in Fig.~\ref{fig4}(b) resolves a finite minimum separation of approximately 0.009~cm$^{-1}$, while $s^{z}$ changes sign across the avoided crossing, indicating exchange of circular-polarization character. A second avoided crossing between the $A_{g}$(2)-derived branches ($\nu = 15, 16$) produces the strongest single-branch chiral response among the low-frequency Raman-active modes [Fig.~\ref{fig4}(c)].

A distinct mechanism appears along T--Z, where the nearly degenerate $B_{1g}$(2) mode pair ($\nu = 13, 14$) remains locked by nonsymmorphic $2_{1}$ screw/glide symmetry [Fig.~\ref{fig4}(d)]. The two partner branches carry equal and opposite $s^{z}$, reflecting opposite circular-polarization character associated with symmetry-enforced band sticking rather than avoided crossing. Similar degenerate pairs are shown in Fig.~S7~\cite{SI}. In contrast, path segments such as Y--$\Gamma$ and U--R show negligible PAM, consistent with symmetry constraints that favor predominantly real phonon eigenvectors and suppress circular lattice motion. Across all modes and momenta, the dominant PAM contribution is $s^{z}$, corresponding to circular motion within the $ac$ plane, consistent with the layered structure of ZrSnS$_3$. These results provide a microscopic picture of helicity-sensitive vibrational response in quasi-one-dimensional ZrSnS$_3$ and highlight how weak symmetry relaxation, mode hybridization, and nonsymmorphic band sticking can generate angular-momentum-like phonon textures in low-symmetry van der Waals semiconductors.

\begin{figure*}[t]
\centering
\includegraphics[width=0.7\textwidth]{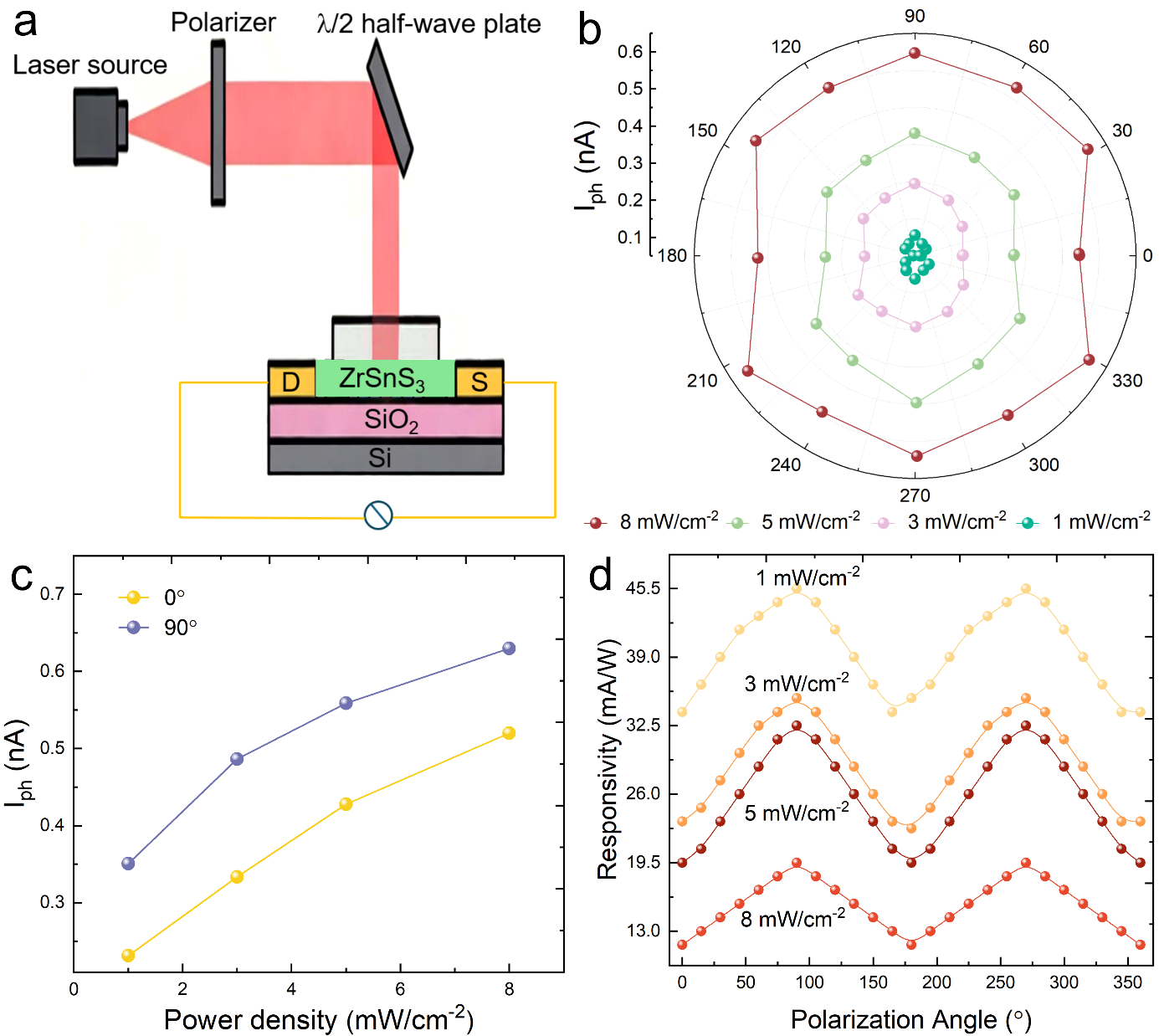}
\caption{\textbf{Polar dependence of photoresponse characteristics under 520~nm illumination.} (a) Schematic diagram measuring the photocurrent of the device under varying polarization light. (b) Polar plot of photocurrent at different light power densities (1, 3, 5, and 8~mW/cm$^2$), showing angular dependence. (c) Photocurrent as a function of light power density at 0$^\circ$ and 90$^\circ$ polarization angles. (d) Responsivity as a function of polarization angle at different light power densities (1, 3, 5, and 8~mW/cm$^2$). A bias of 10~V was used for the measurement.}
\label{fig6}
\end{figure*}

The optoelectronic properties of the ZrSnS$_3$ nanowires were investigated through photodetection measurements, as summarized in Fig.~\ref{fig5}. Figure~\ref{fig5}(a) shows the device structure where a ZrSnS$_3$ nanowire bridges two gold electrodes on a SiO$_2$/Si substrate. Figure~\ref{fig5}(b) confirms the proper alignment of the nanowire, with a clean contact interface and a channel length of approximately 10~$\mu$m. Figures~\ref{fig5}(c--e) display the $I$-$V$ characteristics under different gate voltages ($V_{\mathrm{g}} = 0$, $\pm 60$~V) and light intensities from a 520~nm laser. The symmetric curves indicate ohmic contact, with minimal Schottky barrier effects. Illumination significantly increases the current, confirming the material's high photosensitivity. The photocurrent increases with gate voltage in both directions, attributed to enhanced carrier density in the nanowire channel. The linear increase of photocurrent within a bias range of $-10$~V to $+10$~V reflects the semiconducting nature of ZrSnS$_3$ and effective gate control over carrier concentration.

The photocurrent dependence on light intensity in Fig.~\ref{fig5}(f) follows a power-law relation $I_{\mathrm{ph}} \propto P^{\alpha}$ with an exponent $\alpha = 0.38$, indicating the presence of trap states that capture photogenerated carriers and lead to a sub-linear response and slower carrier dynamics. In Fig.~\ref{fig5}(g), the calculated photoresponsivity reaches 50~mA/W at 1~mW/cm$^2$ under a 10~V bias, but decreases with increasing power density due to trap state filling and increased recombination, a common feature in semiconductor-based photodetectors. The device's stability and transient response, shown in Figs.~\ref{fig5}(h) \& (i), demonstrate excellent reproducibility with minimal photocurrent degradation over repeated ON-OFF cycles. The current drops by 0.11~nA when the power is reduced from 7~mW/cm$^2$ to 3~mW/cm$^2$, indicating a stable device response. The rise time of 80~ms and the fall time of 86~ms suggest fast carrier dynamics, with trapping and detrapping processes not significantly hindering switching speed. ZrSnS$_3$ nanowires offer a balance of sensitivity and fast response, with potential for optimization through surface passivation and improved metal contacts to reduce dark current and enhance responsivity. These results position ZrSnS$_3$ as a promising candidate for stable, gate-tunable visible-light photodetectors in nanoscale optoelectronic systems.

The device was further measured under varying polarization angles and laser power density to investigate the polarization response of the current, as shown in Fig.~\ref{fig6}. The photocurrent varies significantly and exhibits a sinusoidal relationship with the polarization angle. Figure~\ref{fig6}(a) represents the schematic illustration of polarization-dependent photocurrent measurement. Figure~\ref{fig6}(b) shows that the photocurrent increases with increasing power density and exhibits a clear angular dependence, with a stronger response at specific polarization angles. The polar plot exhibits clear maxima at specific orientations (e.g., 0$^\circ$, 90$^\circ$, 180$^\circ$, 270$^\circ$) and minima at intermediate angles, indicating a highly anisotropic behavior of the photocurrent. This angular dependence becomes more pronounced as the light power density increases, with the 8~mW/cm$^2$ condition showing the largest photocurrent values and the most distinct angular variation. In Fig.~\ref{fig6}(c), both 0$^\circ$ and 90$^\circ$ polarization angles show a monotonic increase in $I_{\mathrm{ph}}$ with increasing power density, with the 90$^\circ$ angle consistently yielding a higher photocurrent. At all power densities, the 90$^\circ$ polarization yields a consistently higher photocurrent than 0$^\circ$, confirming that the device's photoresponse is not only anisotropic but also exhibits a preferred polarization direction. Both curves show a monotonic increase in $I_{\mathrm{ph}}$ with power density, indicating that the overall photogeneration efficiency scales with incident light intensity, while the relative difference between the two polarizations remains significant.

Figure~\ref{fig6}(d) further demonstrates that the device's responsivity is also strongly polarization-dependent. At each power density, responsivity displays distinct peaks at specific angles (e.g., $\sim$90$^\circ$ and $\sim$270$^\circ$), mirroring the anisotropy seen in the photocurrent. The magnitude of these peaks increases with power density, with the 1~mW/cm$^2$ condition achieving the highest responsivity values ($\sim$45~mA/W), while the 8~mW/cm$^2$ condition shows the lowest ($\sim$13~mA/W). This confirms that the device's sensitivity to light is not only intensity-dependent but also highly sensitive to the orientation of the incident light's electric field. Such behavior makes the device promising for applications in polarization-sensitive photodetection.

\section{Conclusion}

In summary, our comprehensive study of the quasi-one-dimensional van der Waals semiconductor ZrSnS$_3$ highlights its profound structural anisotropy, which directly influences both its vibrational and optoelectronic properties. Using angle-resolved and temperature-dependent Raman spectroscopy, we reveal significant anisotropic phonon dispersion and anharmonic behavior, driven by phonon-phonon interactions. Notably, we observe chiral phonons, evidenced by helicity-dependent variations in Raman intensity under circularly polarized light, which result from the intrinsic polarization of the zigzag Zr/Sn chains in the crystal structure. The combined experimental observations and phonon circular-polarization analysis establish a connection between helicity-sensitive lattice dynamics and anisotropic optoelectronic functionality in ZrSnS$_3$. Beyond fundamental lattice dynamics, we demonstrate the practical application of these properties in a functional device. A ZrSnS$_3$ nanowire photodetector with a high photoresponsivity of 50~mA/W confirms the material's excellent potential for sensitive optoelectronic applications. These results highlight the device's strong polarization anisotropy, where both photocurrent and responsivity are highly sensitive to the angle of the incident light's polarization. The helicity-dependent Raman response and strong anisotropic photoresponse emphasize the importance of anisotropic light-lattice and carrier-phonon coupling in this system. Therefore, ZrSnS$_3$ emerges as a promising material platform for both fundamental research and technological advancement. Its combination of structural anisotropy, chiral phonons, and strong polarized optoelectronic performance positions it as an ideal candidate for polarization-sensitive photodetectors, directional transport devices, and further exploration of quantum phenomena in low-symmetry, low-dimensional systems.

\section{Experiments and Methods}

\subsection{Single crystal synthesis}

The 1D single crystals of ZrSnS$_3$ were grown by using the chemical vapor transport (CVT) method. The precursor powder materials Zr (99.95\% purity), Sn (99.99\% purity), and S (99.99\% purity) were used for the growth process which was purchased from Sigma-Aldrich. The stoichiometric ratio (1:1:3) of Zr, Sn, and S was mixed along with iodine (I) as a transport agent of a mass ratio (1:10) of the total ampoule. After properly mixing the materials, the materials were then sealed in a 25~cm-long quartz tube of 10~mm diameter, evacuated under a vacuum of $10^{-4}$~Pa. The quartz tube was kept in a two-zone temperature-controlled CVT furnace by setting the temperature of the precursor materials side to 900~$^\circ$C and the other side to 950~$^\circ$C (this was set so that the Sn powder would not fly to the cold zone when the temperature increased). However, after reaching this temperature in 6 hours, the temperature of the powder materials side was set to 950~$^\circ$C (for the reaction zone) and the opposite side to 900~$^\circ$C (growth zone) in 48 hours and sustained this temperature for 7 days, then cooled down naturally to room temperature in 5 hours. Following the proper procedure of the crystallization method, the high-quality crystal grows on the low-temperature side of the quartz tube.

\subsection{Raman Measurements}

\textit{Raman scattering:} Raman scattering experiments were performed using a set of single-frequency lasers: 405~nm, 488~nm, 515~nm, 561~nm, 633~nm, 660~nm, and 785~nm. The sample was mounted on the cold finger of a continuous-flow cryostat, allowing measurements to be conducted at temperatures ranging from 5 to 300~K. The excitation beam was focused onto the sample using a 50$\times$ long-working-distance objective (NA = 0.55), producing a spot size of approximately 1~$\mu$m in diameter. The scattered light was collected in a backscattering geometry through the same objective. The Raman signal was dispersed by a 0.75~m spectrometer (Teledyne Princeton Instruments) equipped with an 1800 grooves/mm grating and detected by a liquid-nitrogen-cooled CCD camera. To suppress the reflected laser light and Rayleigh scattering, an appropriate long-pass filter or a set of Bragg-grating notch filters (785~nm) was employed.

\textit{Linear polarization measurements:} Linear polarization-resolved angle-dependent Raman measurements were performed under 785~nm excitation by placing a $\lambda/2$ plate directly above the objective lens, enabling simultaneous rotation of both the incident and scattered light polarizations with respect to the crystal orientation. The measurements were carried out in two linear polarization configurations: co-polarized (XX, $\bar{y}(zz)y$) and cross-polarized (XY, $\bar{y}(zx)y$), corresponding to the detection polarizer being aligned parallel or perpendicular to the excitation polarizer, respectively. In both cases, the incident and scattered light propagate along the $y$-direction. For the XX configuration, the polarizations of both the incident and scattered light are along the $z$-direction, whereas for the XY configuration, the incident polarization is along $z$ and the scattered polarization is along the perpendicular $x$-direction. Note that the orientations of the $x$-, $y$-, and $z$-directions with respect to the crystallographic axes are given above.

\textit{Helicity measurement:} Circularly polarized and helicity-resolved Raman measurements were performed using fixed linear polarizers in the excitation and detection paths, together with $\lambda/4$ wave plates to control the polarization states. Circular polarization configurations were selected by rotating the $\lambda/4$ wave plates in the excitation and detection paths to realize co-circular ($\sigma^{+}/\sigma^{+}$ and $\sigma^{-}/\sigma^{-}$) and cross-circular ($\sigma^{+}/\sigma^{-}$ and $\sigma^{-}/\sigma^{+}$) geometries. The chirality-dependent Raman response was studied using the following procedure: the excitation light was circularly polarized ($\sigma^{+}$ or $\sigma^{-}$), while the detection was effectively unpolarized. Experimentally, this was achieved by summing the signals collected under the same circular excitation but different detection configurations (e.g., $\sigma^{+}/\sigma^{+}$ and $\sigma^{+}/\sigma^{-}$).

\subsection{Device Fabrication and Measurements}

\textit{Sample Preparation:} High-quality ZrSnS$_3$ flakes were mechanically exfoliated from bulk crystals using white Scotch Magic tape. The exfoliated layers were first transferred onto a polydimethylsiloxane (PDMS) substrate provided by Nanjing Muke Nano Technology Co., Ltd. Flakes with uniform contrast and suitable thickness were identified under an optical microscope. For device fabrication, selected ZrSnS$_3$ layers were sequentially transferred onto pre-patterned gold electrodes, which were defined by standard photolithography followed by electron beam evaporation. The overlap between the flake and electrodes ensured reliable electrical contact for subsequent measurements.

\textit{Electrical Transport Measurements:} Electrical transport measurements of the ZrSnS$_3$ devices were performed using a CRX-VF probe station integrated with an Agilent B1500 semiconductor analyzer. Prior to measurements, the system was evacuated to minimize the influence of adsorbed moisture and oxygen. A two-probe configuration was adopted, where a bias voltage ($V_{\mathrm{ds}}$) was applied across the ZrSnS$_3$ channel, and the corresponding current ($I_{\mathrm{ds}}$) was recorded. The bias voltage was swept between $-10$~V and $+10$~V to evaluate the current--voltage characteristics. Throughout the measurements, the gate voltage ($V_{\mathrm{g}}$) was maintained at 0~V to focus on the intrinsic transport behavior of ZrSnS$_3$. All measurements were conducted at room temperature. For polarized photodetection, a half-wave plate was used to rotate the polarization direction of the lasers.

\subsection{Density functional theory calculations}

\textit{Methods:} First-principles DFT calculations were carried out using Quantum ESPRESSO \cite{Giannozzi2009}. Norm-conserving Vanderbilt pseudopotentials \cite{Hamann2013} generated with the Perdew-Burke-Ernzerhof functional were employed. A plane-wave kinetic-energy cutoff of 80~Ry and a $\Gamma$-centered Monkhorst-Pack $k$-point grid of $4 \times 8 \times 2$ were used to describe the electronic structure. Structural relaxations were carried out until the total energy converged within $10^{-6}$~Ry and the residual forces on each atom were smaller than $10^{-4}$~Ry~\AA$^{-1}$. Phonon properties were calculated using density-functional perturbation theory (DFPT) \cite{Baroni2001}. Dynamical matrices and the linear variation of the self-consistent potential were computed on a $2 \times 4 \times 1$ $\mathbf{q}$-mesh and subsequently Fourier-interpolated to obtain phonon dispersions along high-symmetry directions. Raman-active phonon modes at the $\Gamma$ point and corresponding Raman intensities were computed using QERaman \cite{Hung2024}, enabling direct comparison with experimental Raman spectra.

\textit{Phonon circular-polarization and PAM calculations:} To characterize the circular-polarization character of phonons, we analyzed a phonon pseudo-angular-momentum-like quantity from the DFPT phonon eigenvectors. This quantity is a continuous expectation value derived from the imbalance of right- and left-handed circular components, rather than a discrete symmetry-protected quantum number. The phonon polarization vector $e(\mathbf{q}, \nu)$ contains full information about the atomic displacement patterns for phonon branch $\nu$ at wavevector $\mathbf{q}$. Circular or elliptical atomic motion can be identified by decomposing each eigenvector into left- and right-handed circular components in the plane perpendicular to the crystallographic $c$ axis. For each atom $j$, we introduce a circular basis consisting of right-handed ($|R_{j}\rangle$), left-handed ($|L_{j}\rangle$), and longitudinal ($|Z_{j}\rangle$) components, where the right- and left-handed basis vectors correspond to equal-amplitude oscillations along orthogonal Cartesian directions with a relative phase of $\pm\pi/2$ \cite{Basak2022}. In this framework, ideal circular motion corresponds to equal amplitudes of the two orthogonal components, while deviations from this condition lead to elliptical trajectories.

Each phonon eigenvector can then be expressed as
\begin{equation}
e = \sum_{j}\left( \alpha_{j}^{R} \mid R_{j}\rangle + \alpha_{j}^{L} \mid L_{j}\rangle + \alpha_{j}^{Z} \mid Z_{j}\rangle \right),
\label{eq6}
\end{equation}
where $\alpha_{j}^{R}$, $\alpha_{j}^{L}$, and $\alpha_{j}^{Z}$ are the projection coefficients onto the circular and longitudinal basis states. The phonon pseudo-angular momentum operator along the $z$ direction is defined as the sum of atomic contributions,
\begin{equation}
S_{z} \equiv \sum_{j = 1}^{N}s_{j}^{z} = \sum_{j}\left( \mid R_{j}\rangle\langle R_{j} \mid - \mid L_{j}\rangle\langle L_{j} \mid \right),
\label{eq7}
\end{equation}
and the phonon PAM is obtained as
\begin{equation}
s^{z} = e^{\dagger}S^{z}e = \sum_{j = 1}^{N}s_{j}^{z}\hbar = \sum_{j = 1}^{N}\left( |\alpha_{j}^{R}|^{2} - |\alpha_{j}^{L}|^{2} \right)\hbar,
\label{eq8}
\end{equation}
where $N$ is the number of atoms in the unit cell. The quantity $s_{j}^{z} = |\alpha_{j}^{R}|^{2} - |\alpha_{j}^{L}|^{2}$ represents the contribution of the $j$-th atom to the phonon chirality. A value $|s_{j}^{z}| = 1$ corresponds to ideal circular motion, $s_{j}^{z} = 0$ to purely linear vibration, and intermediate values to elliptical motion.

\section*{Authors Contributions}

Z.M. and S.B.M. contributed equally to this work. Z.M. carried out most of the experiments and analyzed the data. G., G.K., and M.R.M. performed the Raman scattering measurements. W.A. performed the device characterization. O.I. helped in characterization. Z.U.R. grew the single crystal, and S.B.M. performed the DFT calculations. Z.M., S.B.M., W.A., A.R., M.R.M., X.Y., and W.Z. designed the project and co-wrote the manuscript. All authors discussed the results and contributed to the manuscript writing.

\section*{Conflict of Interest}

The authors declare no competing financial interests.

\begin{acknowledgments}
The project was supported by Beijing Natural Science Foundation (No.\ 4232070), the National Natural Science Foundation of China (No.\ T2394475 and 52261145694), and the International Mobility Project (No.\ B16001). M.R.M.\ acknowledges the support from the National Science Centre, Poland (Grant No.\ 2020/37/B/ST3/02311). S.M.\ acknowledges the Texas Advanced Computing Center (TACC) at UT Austin (\url{http://www.tacc.utexas.edu}) for providing computational resources through the ACCESS program under Award No.\ TG-PHY250052.
\end{acknowledgments}

\bibliography{reference}

\end{document}


\title{Supporting Information: \\ Chiral Phonons and Giant Anisotropic Photoresponse in Quasi-1D van der Waals Semiconductor ZrSnS$_3$}

\author{Zahir Muhammad}
\thanks{Z.M. and S.M. contributed equally to this work.}
\affiliation{National Key Laboratory of Spintronics, Hangzhou International Innovation Institute, Beihang University, Hangzhou 311115, P.R. China}

\author{Shashi B. Mishra}
\email{smishra9@binghamton.edu}
\affiliation{Department of Physics, Binghamton University-SUNY, Binghamton, New York 13902, USA}

\author{Gayatri}
\affiliation{University of Warsaw, Faculty of Physics, 02-093 Warsaw, Poland}

\author{Grzegorz Krasucki}
\affiliation{University of Warsaw, Faculty of Physics, 02-093 Warsaw, Poland}

\author{Wajid Ali}
\email{w.ali2@uw.edu.pl}
\affiliation{University of Warsaw, Faculty of Physics, 02-093 Warsaw, Poland}

\author{Katarzyna Olkowska-Pucko}
\affiliation{University of Warsaw, Faculty of Physics, 02-093 Warsaw, Poland}

\author{Obaid Iqbal}
\affiliation{School of Materials Science and Engineering, Institutes of Physical Science and Information Technology, Anhui University, Hefei 230601, P.R. China}

\author{Zia Ur Rehman}
\affiliation{Nanoscale Synthesis \& Research Laboratory, Department of Applied Physics, University of Karachi, Karachi, Pakistan}

\author{Aziz Ur Rahman}
\email{aziz@ustc.edu.cn}
\affiliation{National Key Laboratory of Spintronics, Hangzhou International Innovation Institute, Beihang University, Hangzhou 311115, P.R. China}

\author{Maciej R. Molas}
\affiliation{University of Warsaw, Faculty of Physics, 02-093 Warsaw, Poland}

\author{Lin Xiaoyang}
\email{XYLin@buaa.edu.cn}
\affiliation{National Key Laboratory of Spintronics, Hangzhou International Innovation Institute, Beihang University, Hangzhou 311115, P.R. China}

\author{Weisheng Zhao}
\affiliation{National Key Laboratory of Spintronics, Hangzhou International Innovation Institute, Beihang University, Hangzhou 311115, P.R. China}

\maketitle

\section*{Supporting characterization and Results}

\begin{figure}[htb]
\centering
\includegraphics[width=0.5\textwidth]{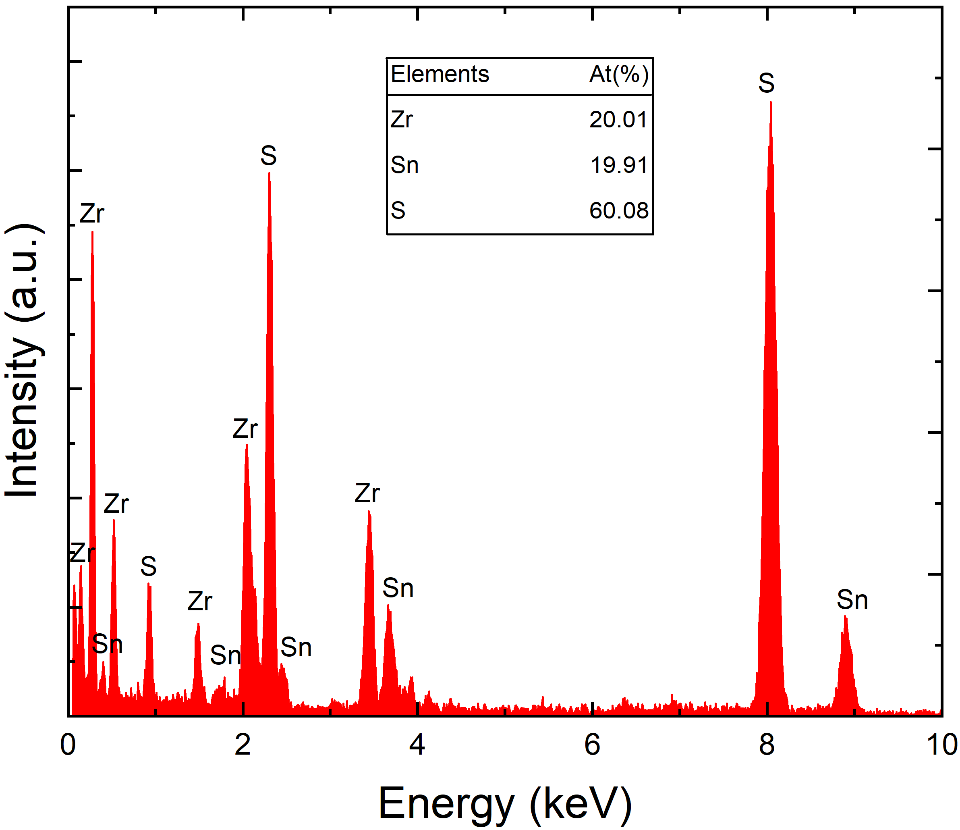}
\caption{EDX spectra of ZrSnS$_3$ single crystal and an inset table of the atomic ratio.}
\label{figS1}
\end{figure}

The EDX spectra show the elemental peaks of Zr, Sn and S, whereas the inset table can identify the stoichiometric ratio of Zr, Sn and S atoms in the crystals. The stoichiometric ratio clearly confirm 1:1:3 of Zr, Sn and S in this single crystal, respectively.

\begin{figure}[htb]
\centering
\includegraphics[width=0.95\textwidth]{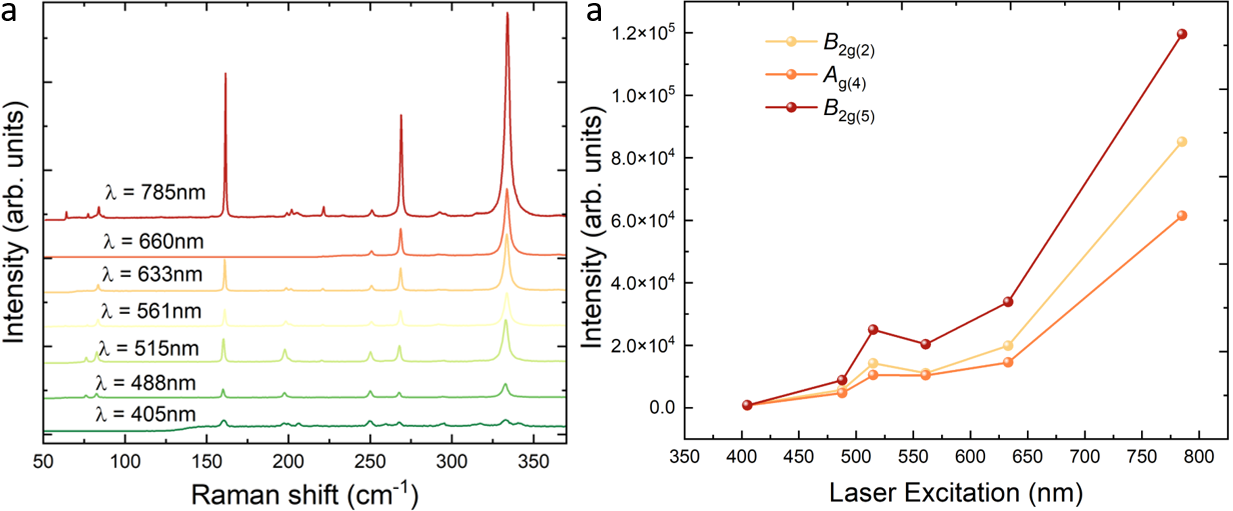}
\caption{(a) Excitation-dependent Raman scattering of the ZrSnS$_3$ and (b) intensity variation of the strongest Raman modes as a function of excitation energy.}
\label{figS2}
\end{figure}

Figure~\ref{figS2}(a) shows the Raman scattering spectra measured for a set of excitation energies. The vibrational modes of ZrSnS$_3$ are strongly enhanced when going from the visible (resonant) regime to the near-infrared (non-resonant) regime, either because the modes are coupled to electronic transitions of the sample or because the optical penetration depth at 785~nm is greater. These are typically modes where the atomic displacements directly contribute to the change in electron distribution during the excitation. The 785~nm laser provided a more comprehensive vibrational profile, whereas the 405~nm laser yielded a less feature-rich spectrum. This may suggest that the 405~nm radiation is strongly absorbed and induces resonance effects that obscure weaker Raman bands; as the excitation wavelength is increased toward the near-infrared, additional modes emerge, and many more Raman modes are resolved at 785~nm, enabling a more representative, non-resonant measurement.

Moreover, we compared the intensities of the three highest peaks that are observed at all the tested excitation energies. The intensities of the peaks are nearly linearly increasing from 405~nm to 785~nm, with a systematic enhancement of almost 10.5 times at 785~nm relative to 405~nm for the strongest peak. This monotonic intensity increase is consistent with a transition from a resonant absorption regime to a transparent, non-resonant regime. At 405~nm, the laser energy likely exceeds the material's band gap, leading to strong absorption and attenuation of the Raman signal. The 785~nm excitation, being sub-bandgap, avoids these effects and thus provides an optimal probe of the intrinsic lattice dynamics.

\section*{Raman modes tensor analysis}

The Raman cross section of a phonon mode that gives the scattered light intensity ($I$) can be written as \cite{Loudon1964}:
\begin{equation}
I \propto \left| \hat{e}_{i} \cdot \hat{R}_{ij} \cdot \hat{e}_{s} \right|^{2}
\label{eqS1}
\end{equation}
here, $\hat{e}_{i} = \left( 0 \;\; \cos\theta \;\; \sin\theta \right)$, while for the scattered light $\hat{e}_{s} = \left( 0 \;\; \cos\theta \;\; \sin\theta \right)$.

\medskip
\noindent\textit{$A_{g}$ modes in the parallel polarization configuration:}

The Raman tensor for the $A_{g}$ modes given in Eq.~(1) of the main text, we can obtain the angular dependence of the intensities of the $A_{g}$ modes:
\begin{gather}
\hat{R}_{A_{g}} = \begin{pmatrix} a & 0 & 0 \\ 0 & b & 0 \\ 0 & 0 & c \end{pmatrix} \nonumber \\
\hat{e}_{i} \cdot \hat{R}_{A_{g}} \cdot \hat{e}_{s} = \begin{pmatrix} e_{i,x} & e_{i,y} & e_{i,z} \end{pmatrix}
\begin{pmatrix} a & 0 & 0 \\ 0 & b & 0 \\ 0 & 0 & c \end{pmatrix}
\begin{pmatrix} e_{s,x} \\ e_{s,y} \\ e_{s,z} \end{pmatrix}
= \left( a e_{i,x}e_{s,x} + b e_{i,y}e_{s,y} + c e_{i,z}e_{s,z} \right) \nonumber \\
I_{A_{g}} \propto \left| a e_{i,x}e_{s,x} + b e_{i,y}e_{s,y} + c e_{i,z}e_{s,z} \right|^{2} \nonumber \\
e_{i} = e_{s} = \hat{x} \;\Longrightarrow\; I \propto |a|^{2} \nonumber \\
e_{i} = e_{s} = \hat{y} \;\Longrightarrow\; I \propto |b|^{2} \nonumber \\
e_{i} = e_{s} = \hat{z} \;\Longrightarrow\; I \propto |c|^{2}
\label{eqS2}
\end{gather}

\medskip
\noindent\textit{$B_{1g}$ modes in the parallel polarization configuration:}
\begin{gather}
\hat{R}_{B_{1g}} = \begin{pmatrix} 0 & d & 0 \\ d & 0 & 0 \\ 0 & 0 & 0 \end{pmatrix} \nonumber \\
\hat{e}_{i} \cdot \hat{R}_{B_{1g}} \cdot \hat{e}_{s} = \begin{pmatrix} e_{i,x} & e_{i,y} & e_{i,z} \end{pmatrix}
\begin{pmatrix} 0 & d & 0 \\ d & 0 & 0 \\ 0 & 0 & 0 \end{pmatrix}
\begin{pmatrix} e_{s,x} \\ e_{s,y} \\ e_{s,z} \end{pmatrix}
= d\left( e_{i,x}e_{s,y} + e_{i,y}e_{s,x} \right) \nonumber \\
I_{B_{1g}} \propto \left| d\left( e_{i,x}e_{s,y} + e_{i,y}e_{s,x} \right) \right|^{2} \nonumber \\
e_{i} = \hat{x},\; e_{s} = \hat{y} \;\Longrightarrow\; I \propto |d|^{2} \nonumber \\
e_{i} = \hat{y},\; e_{s} = \hat{x} \;\Longrightarrow\; I \propto |d|^{2}
\label{eqS3}
\end{gather}

\medskip
\noindent\textit{$B_{2g}$ modes in the parallel polarization configuration:}
\begin{gather}
\hat{R}_{B_{2g}} = \begin{pmatrix} 0 & 0 & e \\ 0 & 0 & 0 \\ e & 0 & 0 \end{pmatrix} \nonumber \\
\hat{e}_{i} \cdot \hat{R}_{B_{2g}} \cdot \hat{e}_{s} = \begin{pmatrix} e_{i,x} & e_{i,y} & e_{i,z} \end{pmatrix}
\begin{pmatrix} 0 & 0 & e \\ 0 & 0 & 0 \\ e & 0 & 0 \end{pmatrix}
\begin{pmatrix} e_{s,x} \\ e_{s,y} \\ e_{s,z} \end{pmatrix}
= e\left( e_{i,x}e_{s,z} + e_{i,z}e_{s,x} \right) \nonumber \\
I_{B_{2g}} \propto \left| e\left( e_{i,x}e_{s,z} + e_{i,z}e_{s,x} \right) \right|^{2} \nonumber \\
e_{i} = \hat{x},\; e_{s} = \hat{z} \;\Longrightarrow\; I \propto |e|^{2} \nonumber \\
e_{i} = \hat{z},\; e_{s} = \hat{x} \;\Longrightarrow\; I \propto |e|^{2}
\label{eqS4}
\end{gather}

\medskip
\noindent\textit{$B_{3g}$ modes in the parallel polarization configuration:}
\begin{gather}
\hat{R}_{B_{3g}} = \begin{pmatrix} 0 & 0 & 0 \\ 0 & 0 & f \\ 0 & f & 0 \end{pmatrix} \nonumber \\
\hat{e}_{i} \cdot \hat{R}_{B_{3g}} \cdot \hat{e}_{s} = \begin{pmatrix} e_{i,x} & e_{i,y} & e_{i,z} \end{pmatrix}
\begin{pmatrix} 0 & 0 & 0 \\ 0 & 0 & f \\ 0 & f & 0 \end{pmatrix}
\begin{pmatrix} e_{s,x} \\ e_{s,y} \\ e_{s,z} \end{pmatrix}
= f\left( e_{i,y}e_{s,z} + e_{i,z}e_{s,y} \right) \nonumber \\
I_{B_{3g}} \propto \left| f\left( e_{i,y}e_{s,z} + e_{i,z}e_{s,y} \right) \right|^{2} \nonumber \\
e_{i} = \hat{y},\; e_{s} = \hat{z} \;\Longrightarrow\; I \propto |f|^{2} \nonumber \\
e_{i} = \hat{z},\; e_{s} = \hat{y} \;\Longrightarrow\; I \propto |f|^{2}
\label{eqS5}
\end{gather}

\begin{figure}[htb]
\centering
\includegraphics[width=0.8\textwidth]{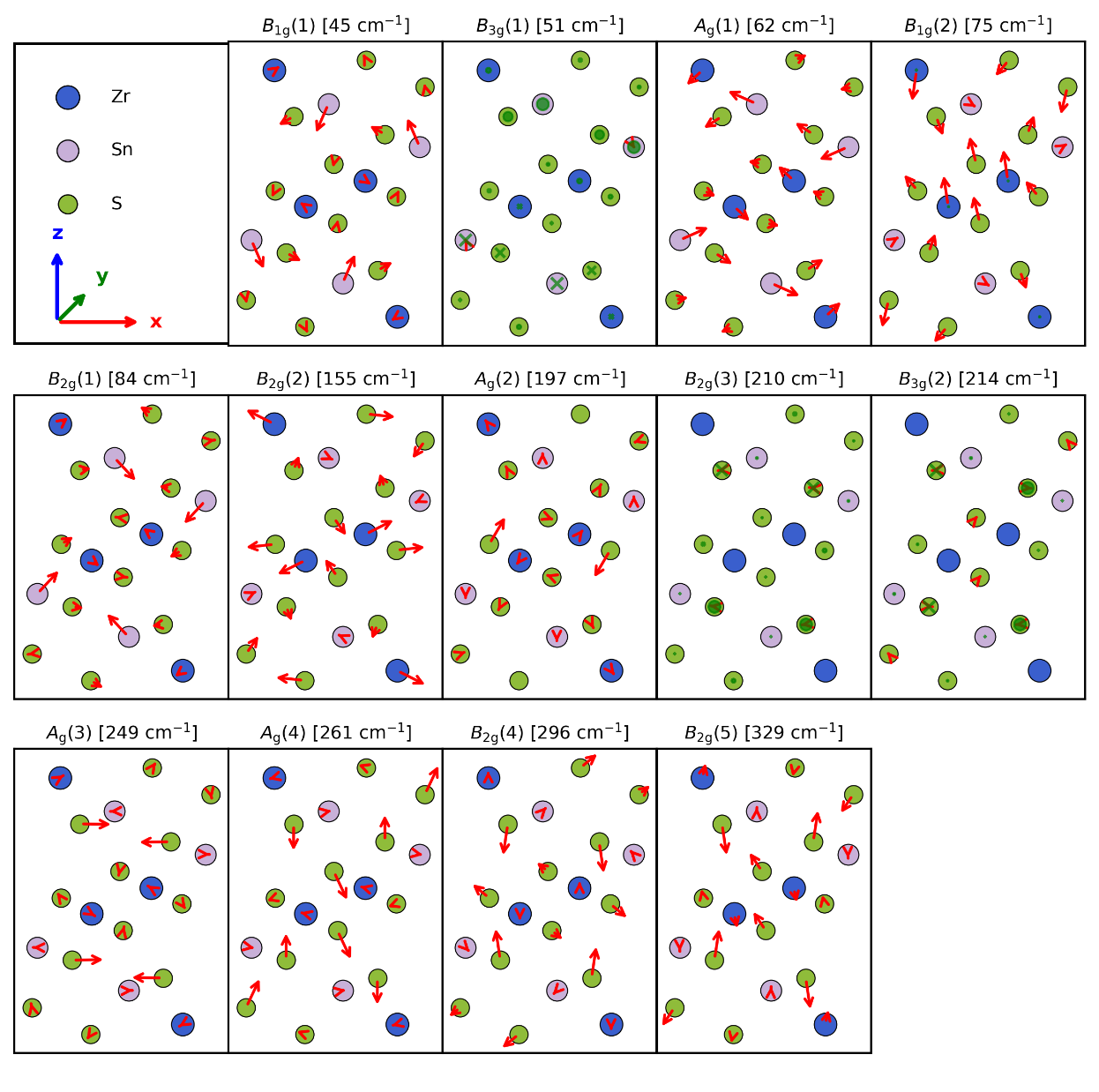}
\caption{Atomic displacements for the Raman modes in ZrSnS$_3$ measured at 5~K using 785~nm (1.52~eV) excitation.}
\label{figS3}
\end{figure}

\begin{table}[htb]
\caption{Raman modes fitting and extracted parameters.}
\label{tabS1}
\begin{ruledtabular}
\begin{tabular}{lcccccc}
Parameters & \multicolumn{6}{c}{Raman Modes} \\
\cline{2-7}
 & $B_{1g}$(1) & $A_{g}$(1) & $B_{2g}$(1) & $B_{2g}$(2) & $A_{g}$(4) & $B_{2g}$(5) \\
\hline
$\omega_{0}$ (cm$^{-1}$) & 44.70 & 61.71 & 81.46 & 159 & 268.5 & 331.5 \\
$A$ (cm$^{-1}$) & $-0.0013$ & $-0.0078$ & $-0.001$ & $-0.0023$ & $-0.0003$ & 0.0007 \\
$B$ (cm$^{-1}$) & 0.00007 & $-0.0002$ & $-0.0006$ & $-0.0001$ & $-0.0002$ & $-0.0003$ \\
$\Gamma_{0}$ (cm$^{-1}$) & 0.5 & 0.3 & 1.13 & 0.69 & 1.33 & 3.31 \\
$C$ (cm$^{-1}$) & 0.0068 & 0.00028 & 0.0052 & 0.0085 & 0.0105 & 0.0226 \\
$D$ (cm$^{-1}$) & 0.0002 & 0.00003 & 0.0001 & 0.00002 & 0.0004 & 0.0002 \\
\end{tabular}
\end{ruledtabular}
\end{table}

\begin{figure}[htb]
\centering
\includegraphics[width=\textwidth]{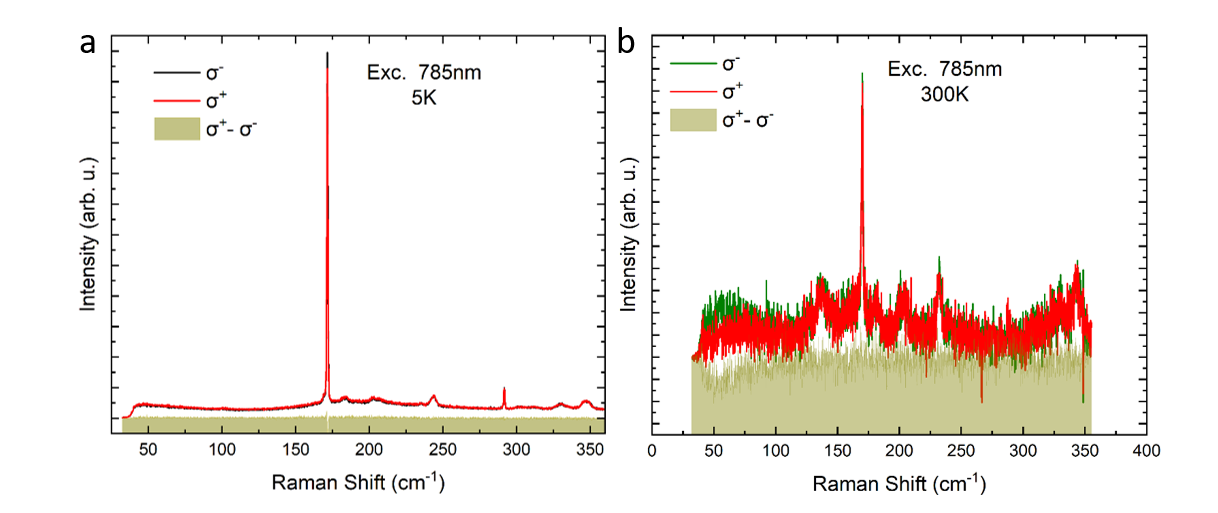}
\caption{Helicity-resolved Raman spectra on bilayer MoTe$_2$. Helicity-resolved Raman spectra under left and right circular polarization ($\sigma^{+}$ and $\sigma^{-}$) and the corresponding difference at (c) 5~K and (d) 300~K.}
\label{figS4}
\end{figure}

Figure~\ref{figS4} presents the controlled measurement on bilayer MoTe$_2$ for comparison using the same helicity-resolved Raman spectra at 5~K and 300~K. The observed results clearly indicate that there is no difference between the opposite helicities in the Raman intensity at both temperatures, confirming the validity of the chiral phonon in ZrSnS$_3$.

\begin{figure}[htb]
\centering
\includegraphics[width=\textwidth]{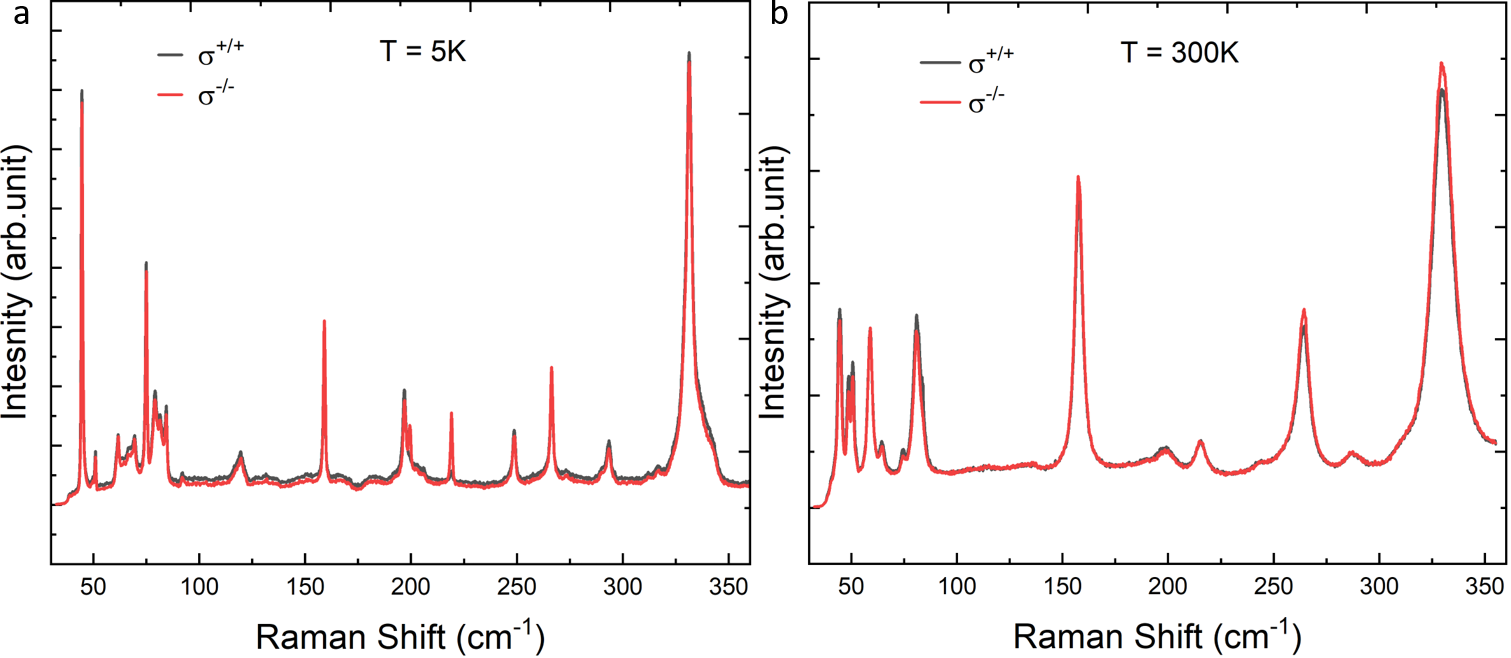}
\caption{The phonon intensities comparison of the helicity-conserving scattering configurations ($\sigma^{+}/\sigma^{+}$ and $\sigma^{-}/\sigma^{-}$) at (a) 5~K and (b) 300~K.}
\label{figS5}
\end{figure}


\begin{table}[!htb]
\caption{Calculated phonon angular-momentum expectation values $s^{\alpha}(\Gamma)$ for all Raman-active modes of ZrSnS$_3$. All components are zero, indicating that $\Gamma$-point phonons are strictly non-chiral and possess real eigenvectors.}
\label{tabS2}
\begin{ruledtabular}
\begin{tabular}{lccccc}
Irrep & Mode\# & Frequency (cm$^{-1}$) & $s^{x}(\Gamma)$ & $s^{y}(\Gamma)$ & $s^{z}(\Gamma)$ \\
\hline
$B_{1g}$(1) & 7  & 50.27  & 0.00 & 0.00 & 0.00 \\
$A_{g}$(1)  & 8  & 51.46  & 0.00 & 0.00 & 0.00 \\
$B_{3g}$(1) & 11 & 62.17  & 0.00 & 0.00 & 0.00 \\
$B_{1g}$(2) & 13 & 64.69  & 0.00 & 0.00 & 0.00 \\
$A_{g}$(2)  & 15 & 68.52  & 0.00 & 0.00 & 0.00 \\
$A_{g}$(3)  & 19 & 78.14  & 0.00 & 0.00 & 0.00 \\
$A_{g}$(4)  & 22 & 115.70 & 0.00 & 0.00 & 0.00 \\
$B_{1g}$(3) & 23 & 152.73 & 0.00 & 0.00 & 0.00 \\
$A_{g}$(5)  & 25 & 163.95 & 0.00 & 0.00 & 0.00 \\
$B_{1g}$(4) & 29 & 186.64 & 0.00 & 0.00 & 0.00 \\
$A_{g}$(6)  & 33 & 209.58 & 0.00 & 0.00 & 0.00 \\
$B_{1g}$(5) & 37 & 227.69 & 0.00 & 0.00 & 0.00 \\
$A_{g}$(7)  & 42 & 243.13 & 0.00 & 0.00 & 0.00 \\
\end{tabular}
\end{ruledtabular}
\end{table}

\begin{table}[htb]
\caption{Raman-active phonon frequencies in the centrosymmetric $Pnma$ and polar $Pna2_{1}$ structures of ZrSnS$_3$. The frequency shift $\Delta\omega = \omega(Pna2_{1}) - \omega(Pnma)$ is given for all 30 Raman-active modes. Mean $|\Delta\omega| = 0.81$~cm$^{-1}$, max $|\Delta\omega| = 7.50$~cm$^{-1}$, RMS $\Delta\omega = 1.63$~cm$^{-1}$. The single outlier ($*$) arises from $B_{1g}$(1)/$B_{2g}$(2) hybridization through the small polar shift in the low-$\omega$, dense-mode region; all 13 experimentally observable Raman peaks ($>$100~cm$^{-1}$) shift by $\leq$1~cm$^{-1}$.}
\label{tabS3}
\begin{ruledtabular}
\begin{tabular}{lcccc}
Label & $Pnma$ \# & $\omega(Pnma)$ cm$^{-1}$ & $\omega(Pna2_{1})$ cm$^{-1}$ & $\Delta\omega$ cm$^{-1}$ \\
\hline
$A_{g}$(1)  & 7  & 50.60  & 50.30  & $-0.30$ \\
$A_{g}$(2)  & 12 & 64.10  & 63.90  & $-0.20$ \\
$A_{g}$(3)  & 16 & 73.00  & 73.20  & $+0.20$ \\
$A_{g}$(4)  & 23 & 152.90 & 152.70 & $-0.20$ \\
$A_{g}$(5)  & 31 & 188.20 & 187.60 & $-0.60$ \\
$A_{g}$(6)  & 41 & 240.70 & 240.90 & $+0.20$ \\
$A_{g}$(7)  & 46 & 256.60 & 256.90 & $+0.30$ \\
$A_{g}$(8)  & 51 & 284.00 & 283.70 & $-0.30$ \\
$A_{g}$(9)  & 54 & 302.70 & 302.50 & $-0.20$ \\
$A_{g}$(10) & 58 & 318.30 & 318.10 & $-0.20$ \\
$B_{1g}$(1) & 9  & 59.20  & 51.70  & $-7.50$ $*$ \\
$B_{1g}$(2) & 18 & 75.60  & 76.70  & $+1.10$ \\
$B_{1g}$(3) & 30 & 187.10 & 186.70 & $-0.40$ \\
$B_{1g}$(4) & 34 & 209.70 & 210.00 & $+0.30$ \\
$B_{1g}$(5) & 42 & 244.00 & 243.10 & $-0.90$ \\
$B_{2g}$(1) & 8  & 54.00  & 51.50  & $-2.50$ \\
$B_{2g}$(2) & 13 & 66.80  & 64.70  & $-2.10$ \\
$B_{2g}$(3) & 20 & 85.10  & 85.40  & $+0.30$ \\
$B_{2g}$(4) & 24 & 155.30 & 155.20 & $-0.10$ \\
$B_{2g}$(5) & 32 & 197.10 & 197.20 & $+0.10$ \\
$B_{2g}$(6) & 44 & 248.50 & 249.00 & $+0.50$ \\
$B_{2g}$(7) & 45 & 255.40 & 255.20 & $-0.20$ \\
$B_{2g}$(8) & 52 & 295.80 & 295.60 & $-0.20$ \\
$B_{2g}$(9) & 55 & 309.40 & 309.40 & $0.00$ \\
$B_{2g}$(10)& 59 & 325.50 & 325.30 & $-0.20$ \\
$B_{3g}$(1) & 11 & 61.80  & 60.40  & $-1.40$ \\
$B_{3g}$(2) & 19 & 75.70  & 78.10  & $+2.40$ \\
$B_{3g}$(3) & 29 & 185.90 & 186.60 & $+0.70$ \\
$B_{3g}$(4) & 36 & 213.70 & 213.60 & $-0.10$ \\
$B_{3g}$(5) & 43 & 244.00 & 243.30 & $-0.70$ \\
\end{tabular}
\end{ruledtabular}
\end{table}

\begin{table}[htb]
\caption{Per-atom comparison between the $Pna2_{1}$ and the $Pnma$ structure of ZrSnS$_3$. Both structures use the same lattice parameters, $a = 9.2335$~\AA, $b = 3.7295$~\AA, $c = 13.7789$~\AA. All atoms occupy Wyckoff position $4c$ in the $Pnma$ parent ($x$, $\tfrac{1}{4}$, $z$ orbit). The Zr and S sublattices coincide with the centrosymmetric reference to the precision of the relaxation; only the four Sn atoms shift coherently along the polar $\mathbf{b}$-axis by $\Delta y = +0.003$ ($\mathbf{b} = +0.011$~\AA). Displacements along $\mathbf{a}$ and $\mathbf{c}$ are zero for every atom.}
\label{tabS4}
\begin{ruledtabular}
\begin{tabular}{cccccccc}
\# & Atom & $x$ & $y_{Pnma}$ & $y_{Pna2_{1}}$ & $\Delta y$ (frac) & $\Delta y$ (\AA) & $z$ \\
\hline
1  & S  & 0.33135 & 0.25000 & 0.25000 & 0.00000 & $+0.0000$ & 0.01086 \\
2  & S  & 0.01006 & 0.75000 & 0.75000 & 0.00000 & $+0.0000$ & 0.10860 \\
3  & S  & 0.72936 & 0.25000 & 0.25000 & 0.00000 & $+0.0000$ & 0.21716 \\
4  & S  & 0.22936 & 0.25000 & 0.25000 & 0.00000 & $+0.0000$ & 0.28284 \\
5  & S  & 0.51006 & 0.75000 & 0.75000 & 0.00000 & $+0.0000$ & 0.39140 \\
6  & S  & 0.83135 & 0.25000 & 0.25000 & 0.00000 & $+0.0000$ & 0.48914 \\
7  & S  & 0.16865 & 0.75000 & 0.75000 & 0.00000 & $+0.0000$ & 0.51086 \\
8  & S  & 0.48994 & 0.25000 & 0.25000 & 0.00000 & $+0.0000$ & 0.60860 \\
9  & S  & 0.77064 & 0.75000 & 0.75000 & 0.00000 & $+0.0000$ & 0.71716 \\
10 & S  & 0.27064 & 0.75000 & 0.75000 & 0.00000 & $+0.0000$ & 0.78284 \\
11 & S  & 0.98994 & 0.25000 & 0.25000 & 0.00000 & $+0.0000$ & 0.89140 \\
12 & S  & 0.66865 & 0.75000 & 0.75000 & 0.00000 & $+0.0000$ & 0.98914 \\
13 & Sn & 0.53901 & 0.75000 & 0.75300 & $+0.00300$ & $+0.0112$ & 0.17104 \\
14 & Sn & 0.03901 & 0.75000 & 0.75300 & $+0.00300$ & $+0.0112$ & 0.32896 \\
15 & Sn & 0.96099 & 0.25000 & 0.25300 & $+0.00300$ & $+0.0112$ & 0.67104 \\
16 & Sn & 0.46099 & 0.25000 & 0.25300 & $+0.00300$ & $+0.0112$ & 0.82896 \\
17 & Zr & 0.83654 & 0.25000 & 0.25000 & 0.00000 & $+0.0000$ & 0.04742 \\
18 & Zr & 0.33654 & 0.25000 & 0.25000 & 0.00000 & $+0.0000$ & 0.45258 \\
19 & Zr & 0.66346 & 0.75000 & 0.75000 & 0.00000 & $+0.0000$ & 0.54742 \\
20 & Zr & 0.16346 & 0.75000 & 0.75000 & 0.00000 & $+0.0000$ & 0.95258 \\
\end{tabular}
\end{ruledtabular}
\end{table}


\begin{figure}[!htb]
\centering
\includegraphics[width=\textwidth]{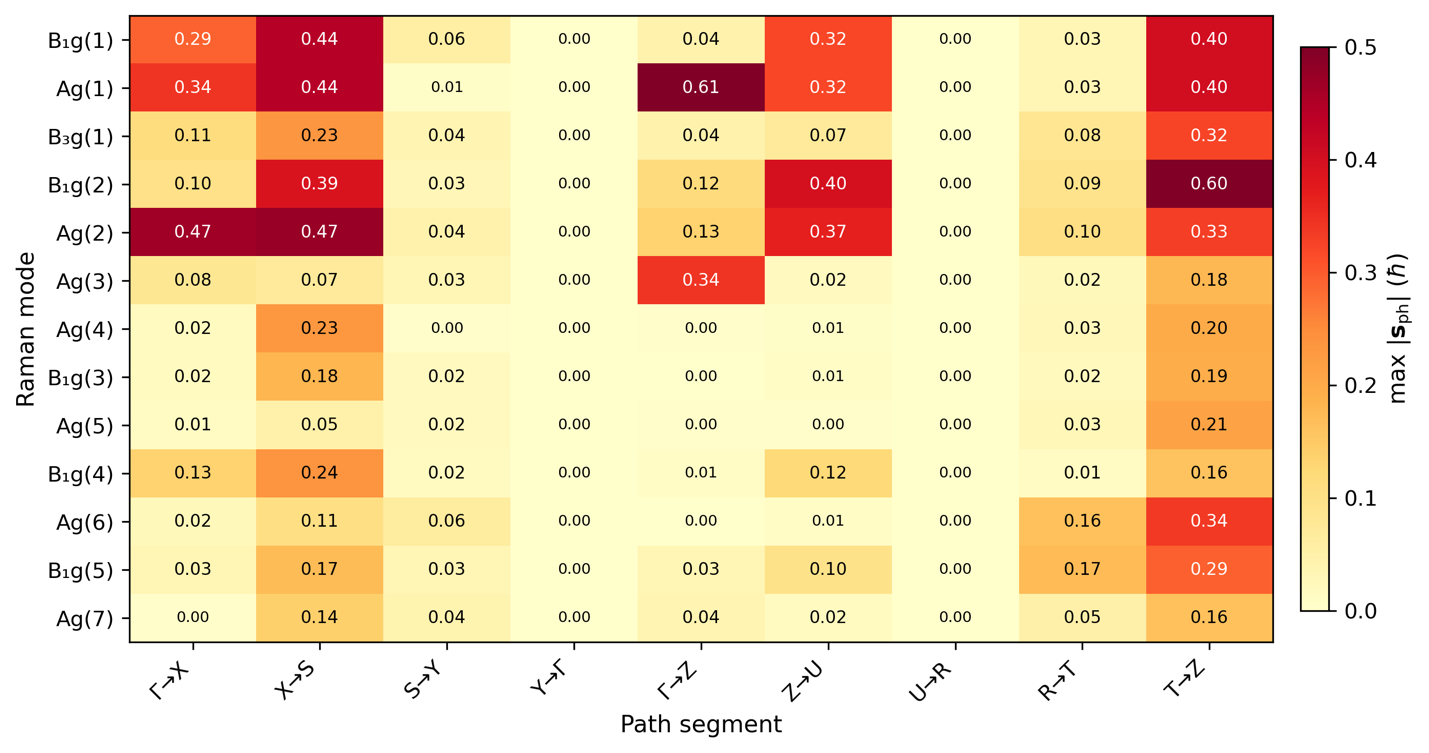}
\caption{Mode- and momentum-resolved phonon chirality in ZrSnS$_3$. Heatmap showing the maximum gauge-invariant phonon pseudo-angular momentum $|s^{z}|$ (in units of $\hbar$) for each Raman-active phonon mode along individual high-symmetry path segments in the Brillouin zone. The color scale quantifies the strength of circular lattice motion associated with each mode-momentum combination.}
\label{figS6}
\end{figure}

\begin{figure}[htb]
\centering
\includegraphics[width=\textwidth]{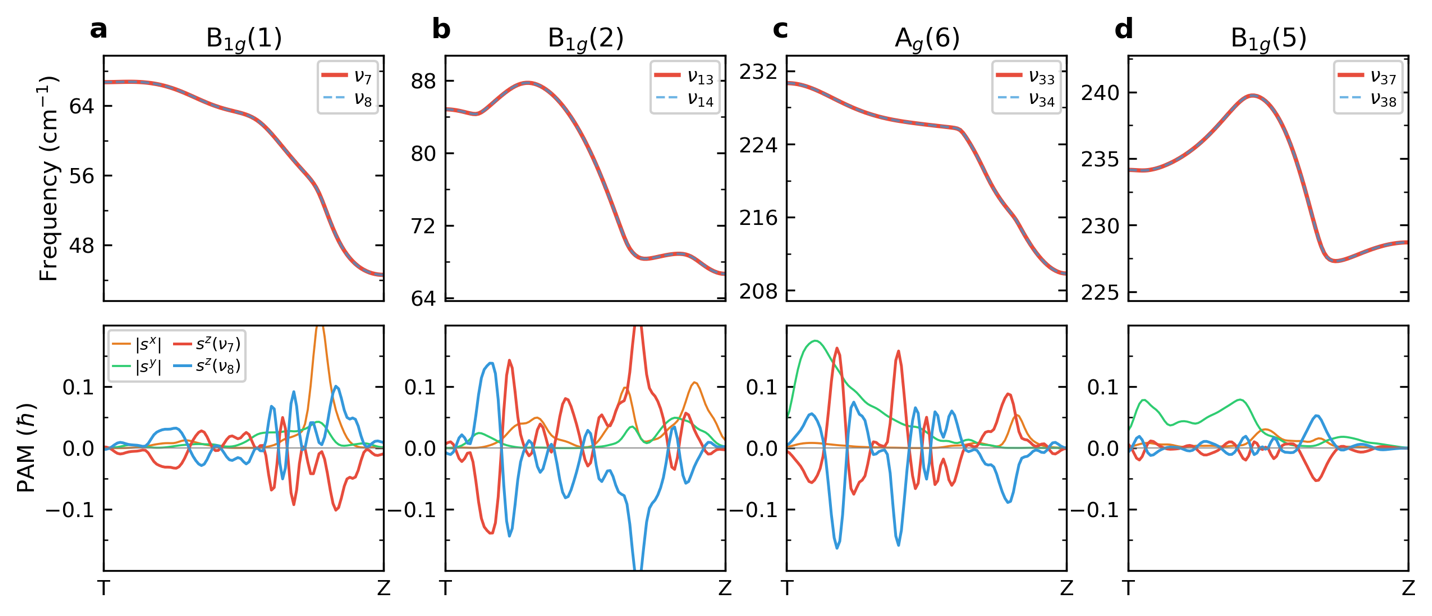}
\caption{Symmetry-protected phonon degeneracies and opposite circular polarization along T--Z. Four representative degenerate phonon pairs along the T--Z direction: (a) $B_{1g}$(1), (b) $B_{1g}$(2), (c) $A_{g}$(6), and (d) $B_{1g}$(5). The upper panels show the nearly identical phonon dispersions of each degenerate pair, while the lower panels show the corresponding PAM components. The two partner modes carry opposite $s^{z}$ components, whereas the gauge-invariant magnitudes $|s^{x}|$ and $|s^{y}|$ remain smooth along the path.}
\label{figS7}
\end{figure}

\bibliography{reference}